\documentclass[10pt,journal,a4]{IEEEtran}

\usepackage[square,sort,comma,numbers]{natbib}
\usepackage{amsmath,amssymb,amsfonts}
\usepackage{algorithm}
\usepackage{algorithmicx}
\usepackage[noend]{algpseudocode}
\usepackage{graphicx}
\usepackage{textcomp}
\usepackage{bbm}
\usepackage{mathabx}
\usepackage{xcolor}
\usepackage[normalem]{ulem}

\newtheorem{theorem}{Theorem}

\def\argmax{\mathop{\rm argmax}}
\def\argmin{\mathop{\rm argmin}}

\newcommand{\sphere}{\ensuremath{\mathbb{S}^2}}

\def\sl{\mathop{\mathfrak s\mathfrak l}\nolimits}

\def \S {\mathbb{S}}

\begin{document}
\title{Optimized Diffusion Imaging for Brain Structural Connectome Analysis}
\author{William Consagra, Arun Venkataraman, and Zhengwu Zhang
\thanks{William Consagra is with the Department of Biostatistics and Computational Biology, University of Rochester, Rochester, NY, 14642 (email: william\_consagra@urmc.rochester.edu). Arun Venkataraman is with the Department of Physics, University of Rochester, Rochester, NY,  14627 (e-mail: arun\_venkataraman@urmc.rochester.edu). Zhengwu Zhang is with the Department of Statistics and Opertaions Research, UNC Chapel Hill, Chapel Hill, NC, 27599 (e-mail: zhengwu\_zhang@unc.edu).}
}

\maketitle

\begin{abstract}
High angular resolution diffusion imaging (HARDI) is a type of diffusion magnetic resonance imaging (dMRI) that measures diffusion signals on a sphere in q-space. It has been widely used in data acquisition for human brain structural connectome analysis. To more accurately estimate the structural connectome,  dense samples in q-space are often acquired, potentially resulting in long scanning times and logistical challenges. This paper proposes a statistical method to select q-space directions optimally and estimate the local diffusion function from sparse observations.  The proposed approach leverages relevant historical dMRI data to calculate a prior distribution to characterize local diffusion variability in each voxel in a template space.  For a new subject to be scanned, the priors are mapped into the subject-specific coordinate and used to help select the best q-space samples.  Simulation studies demonstrate big advantages over the existing HARDI sampling and analysis framework. We also applied the proposed method to the Human Connectome Project data and a dataset of aging adults with mild cognitive impairment. The results indicate that with very few q-space samples (e.g., 15 or 20), we can recover structural brain networks comparable to the ones estimated from 60 or more diffusion directions with the existing methods. 
\end{abstract}


\section{Introduction}

Diffusion Magnetic Resonance Imaging (dMRI) is an advanced MRI methodology that measures the probabilistic motion of water molecules in order to probe tissue microstructure \cite{stejskal}. The earliest applications of dMRI focus on studying the anisotropic diffusion of water in the white matter (WM) using the diffusion tensor imaging (DTI) model \cite{Basser_diffusion}. However, DTI has significant limitations, since it can only describe one major fiber direction within each voxel \cite{tuch2002high}, presenting a significant obstacle for efforts to trace WM pathways in crossing fiber regions \cite{maier2017challenge}. To overcome this limitation, advanced acquisition methods, such as high angular resolution diffusion imaging
(HARDI) \cite{alexander2002detection}, have been proposed to better characterize the underlying WM structure by estimating an object known as the diffusion orientation distribution function (ODF), denoted here as $f$, or its sharper version, the fiber ODF (fODF) obtained by deconvolving the ODF  \cite{descoteaux_2009}. To estimate the structural connectome, tractography algorithms \cite{Tournier2012g,Girard2014g} are applied to the estimated fODF field to compute streamlines connecting different brain regions. Therefore, accurate estimation of the fODF is paramount for performing quality structural connectome analysis.  
\par 
With HARDI data, the ODF or fODF can be estimated from the diffusion signal using the Funk-Radon transform (FRT) \cite{tuch_2004}, making it the most popular dMRI acquisition framework for structural connectome analysis. Commonly, HARDI data are acquired using uniformly distributed diffusion gradient vectors densely sampled on spheres, known as shells, which inhabit q-space. The distribution of these gradient vectors has been a subject of study. Frequently an electrostatic repulsion model is used to generate them \cite{jones1999}. As shown in this paper, such gradient setting can be sub-optimal with respect to recovering the diffusion signal and fODF. Moreover, in practice, scans are often time limited and, consequently, parameters are altered for fast acquisition, e.g., only $15$ or less directions are sampled in some studies \cite{chen2013robust,chang2015human}. Therefore, there are important questions to consider in HARDI: {\bf if we are restricted to a limited number of gradient directions to sample, where in q-space should we sample, and given the sparse samples, how do we recover the fODF for connectome analysis?}  In this paper we address these questions by (a) optimizing the diffusion sampling scheme for  each individual's brain, and (b) estimating diffusion signal from sparse samples. 
\par 
 The methods proposed in this paper leverage the assumption that high-quality dMRI data are available for the population of interest, e.g. healthy subjects in a certain age range. For simplicity, we only consider diffusion signal on a single shell. The diffusion signal at the $v$-th voxel can be considered as a realization of a \textit{random function} ${X}_v \in \mathbb{L}^2(\mathbb{S}^2$), i.e., a measurable function with realizations in the space of square integrable functions over the 2-sphere. In particular, we assume the distribution of ${X}_v$ can be modeled as a Gaussian process. Note that Gaussian process models have previously been used to model the diffusion signal in dMRI \cite{andersson2015}. Employing an empirical Bayesian approach, the historical dMRI data is used to estimate the mean and covariance functions defining ${X}_v$, characterizing the prior distribution of the  diffusion signal across the brain. This prior is used to derive an estimator for a new subject's diffusion signal in a manner that is robust even in the sparsely sampled case. Additionally, we propose a greedy diffusion direction selection algorithm to sequentially select the best directions to acquire dMRI data. The  proposed method shows significant advantages over the current dMRI methods. In particular, we demonstrate that our method can produce a much better structural connectome reconstruction with sparse diffusion samples.
\par 
Incorporating prior information in the design of q-space sampling has been investigated before. For example, \cite{caruyer2009} proposes an optimal sampling scheme in the context of DTI and historical data are used to estimate the probability density function of the diffusion tensor field. \cite{poot2010} uses diffusion kurtosis imaging (DKI) to form a parametric approximation of the diffusion signal and prior information is incorporated via the DKI parameters estimated from historical subjects. These methods are derived in the context of finite dimensional parametric models for diffusion signal estimation (i.e., DTI and DKI) and can be problematic for connectome analysis, owing to their assumptions of the underlying diffusion process. 
\par 
In our framework, we consider the diffusion signal as a random function in $\mathbb{L}^2(\sphere)$ which is capable of representing a far richer class of local diffusion patterns than DTI or DKI.  Accordingly, the optimal q-space acquisition problem for the reconstruction of the diffusion signal at a particular voxel is best formulated as a problem of optimal experimental design for sampling and recovering functional data in  $\mathbb{L}^2(\sphere)$. Although there are a few existing papers concerning optimal experimental design for functional data, most of them deal with functions on $\mathbb{R}$ \cite{ji2017,wu2017}. Our approach makes repeated use of special properties of $\mathbb{L}^2(\sphere)$, especially to incorporate relevant prior information, thus distinguishing our problem from these existing ones. We summarize our contributions as follows:
\begin{enumerate}
    \item The diffusion signal is modeled as a random function in $\mathbb{L}^2(\sphere)$, allowing for more flexibility in the types of signals that can be accommodated than do parametric models such as DTI and DKI. This is particularly important for modeling complex diffusion signals in fiber crossing regions. To our knowledge, this is the first work for optimal design in dMRI which both incorporates historical data for constructing priors and does not induce a fixed finite dimensional parametric model for the diffusion signal.
    \item The proposed method uses historical data (e.g., data from large brain imaging studies  such as the Human Connectome Project (HCP) \cite{Glasser2016ag}) to construct a spatially varying prior distribution of diffusion signals over all voxels in a template space. These template space priors can be ported to the coordinate space of a new subject of interest to facilitate real-time optimal sampling design and sparse sample signal reconstruction, amongst other dMRI tasks of interest. 
    \item We propose a computationally efficient approximate optimal design algorithm that avoids costly high-dimensional optimization or stochastic techniques such as simulated annealing. Further, we give a theoretical performance bound between our approximate solution and the true optimal design.
\end{enumerate}

\par 
The rest of the paper is organized as follows. In Section \ref{sec:methods} we present the proposed methodology. Section \ref{sec:simulation} shows the results of a simulation study comparing the performance of our method with existing ones.  In Section \ref{sec:real_data}, we show real data examples for several tasks related to dMRI processing and structural connectome analyses. Conclusions and future directions are 
presented in Section \ref{sec:conclusion}.

\section{Methods}\label{sec:methods}

\subsection{Statistical Model and ODF Estimation}

Under the HARDI framework, for any subject, the diffusion signal at voxel $v$ at q-space direction $p_m\in\S^2$ is modeled as a noisy observation from a voxel specific random function 
\begin{equation}\label{eqn:statistical_model}
    S_v(p_m) = X_v(p_m) + \epsilon_{vm}
\end{equation}
where the random function $X_v$ is assumed to be a Gaussian process (GP) with voxel specific mean and covariance function $\mu_v(p):=\mathbb{E}[X_v(p)]$ and $C_v(p,t):=\mathbb{E}[(X_v(p)-\mu_v(p))(X_v(p)-\mu_v(t))]$, respectively, which we will denote as $\text{GP}(\mu_v, C_v)$. $X_v$ describes the distribution of diffusion signal in the local brain region $v$ for a population of interest, e.g. healthy adults from age 20-35. The measurement error $\epsilon_{vm}\sim \mathcal{N}(0, \sigma^2)$ models the noise introduced by the scanner and is assumed to be independent of diffusion direction $p_m$ and brain region $v$. Although measurement error in MRI generally follows a Rician distribution, it is not uncommon to use GP models to model the diffusion signal \cite{andersson2015}. In this work, we consider the case of single shell acquisition, i.e., all diffusion data is collected using a single b-value. For notational clarity, we denote $S_v(p_{m}) = S_{vm}$.
\par 
The primary object of interest in tractography, the ODF, can be computed from the diffusion signal on a single sphere in q-space by the Funk-Radon transform (FRT):
$$
f_v(p) = \mathcal{G}(X)({p}) = \int_{r \in \S^2} \delta({r}^\intercal{p})X_v({r})d {r}
$$
where $ \mathcal{G}$ is the FRT, $\delta$ is the Dirac delta function and $f_v$ is the ODF
\cite{tuch_2004}. The authors of \cite{descoteaux_2007} show the Funk-Hecke theorem (FKT) implies that the ODF can be constructed by a simple linear transformation applied to the coefficients of the basis expansion of the diffusion signal with respect to a set of modified spherical harmonic functions. As this representation has a tendency to be overly smooth, the FKT can be applied again to obtain a linear transformation representing a deconvolution operator to obtain the fODF \cite{descoteaux_2009}. Therefore, we focus on optimizing the diffusion signal estimation with respect to the spherical harmonics, denoted as $\{\phi_1,...,\phi_J\}$, as this will result in optimal (f)ODF reconstruction, owing to their simple linear relationship.

\subsection{Sparse Sample Diffusion Signal Estimation}\label{eqn:sparse_sample_estimator}

In this work, we are concerned with selecting the optimal q-space directions for the estimation of the diffusion signal function under a sparsity constraint on the total number of samples, i.e., we collect data in the form $\{(p_{1},S_{v1}), ..., (p_{M}, S_{vM}))\}$ with $M$ as a small integer. Let the vector $\boldsymbol{s}_M=(S_{v1},...,S_{vM})^\intercal$ denote the observed diffusion signals and let $\boldsymbol{P}_M = (p_1, ..., p_M)^\intercal$ be the corresponding gradient directions in q-space. 
\subsubsection{Optimal Reduced Rank Basis}
As discussed, to facilitate future connectome extraction it is important to obtain a continuous representation of the estimated diffusion function with respect to basis functions $\{\phi_1,...,\phi_J\}$. However, using the standard regularized least squares based estimation of the coefficients of $\phi_j$, e.g. from \cite{descoteaux_2007}, with sparse and irregular observations can be problematic, as illustrated in \cite{james2000principal} where a reduced rank solution is promoted. Therefore, we propose a low rank estimator which is robust to the case of sparse samples.
\par 
By the Karhunen-Lo\'eve theorem, the optimal rank $K$ basis system for representing the mean centered random function $X_v - \mu_v$, in terms of minimum expected integrated squared error, is given by the first $K$ eigenfunctions of the covariance function $C_v$, denoted as $\{\psi_{v1},...,\psi_{vK}\}$, which are defined by the integral equation 
$$
    \int_{\S^2} C_v(p,t)\psi_{vk}(t)dt = \rho_{vk}\psi_{vk}(p) \text{ for } k=1,2,..., \infty.
$$
where $\rho_{vk}\ge 0$ is the eigenvalue associated with the $k$'th eigenfunction $\psi_{vk}(p)$. Due to this optimality property, the eigenfunctions are a good candidate for sparse sample modeling. Furthermore, if we impose the restriction that $\psi_{vk}$ lies in the linear span of  $\{\phi_1,...,\phi_J\}$, then there exists a simple linear relationship between the $\psi_{vk}$'s and $\phi_j$'s, and thus we can still leverage the analytic FRT for ODF estimation. This restriction translates the infinite dimensional integral equation above to the following finite dimensional eigenvalue problem:
$$
   \boldsymbol{\Sigma}_v\boldsymbol{b}_{vk} = \rho_{vk} \boldsymbol{b}_{vk},
    \text{ for } k=1,2,..., J, 
$$
with orthonormality restriction $\boldsymbol{b}_{vi}^\intercal\boldsymbol{b}_{vj} =\delta_{ij}$, where the eigenfunctions are approximated by $\psi_{vk}(p) =\sum_{j=1}^J {b}_{vkj}\phi_j(p)$ ($b_{vkj}$ is the $j$-th element in $\boldsymbol{b}_{vk}$) and $\boldsymbol{\Sigma}_v$ is the covariance matrix defined according to
\begin{equation}
\label{eqn:sigma}
\begin{aligned}
    C_v(p,t) &= \sum_{k=1}^\infty \rho_{vk}\psi_{vk}(p)\psi_{vk}(t) \approx \sum_{k=1}^K\rho_{vk}\psi_{vk}(p)\psi_{vk}(t) \\
    &= \sum_{k=1}^K \rho_{vk}(\sum_{j=1}^J \boldsymbol{b}_{vk,j}\phi_j(p))(\sum_{l=1}^J \boldsymbol{b}_{vk,l}\phi_l(t)) \\
    &\equiv \boldsymbol{\phi}(p)^\intercal\boldsymbol{\Sigma}_v\boldsymbol{\phi}(t),
\end{aligned}
\end{equation}
where $\boldsymbol{\phi}(p) = (\phi_1(p), ..., \phi_J(p))^\intercal$. 
\subsubsection{Sparse Estimator}
We now present the proposed estimator for the diffusion signal function given sparse q-space samples. For notational convenience, we temporarily drop the subscript $v$, with the understanding that the estimates are formed per voxel. From the discussion above, we would like to obtain an estimate of $X$ in the form $X(p) = \mu(p) + \sum_{k=1}^K \xi_k \psi_k(p)$, where $\xi_k = \int_{\mathbb{S}^{2}} (X(p)-\mu(p)) \psi_k(p)dp$ are the coefficients of the diffusion signal associated with the $k$'th eigenfunction. Due to the assumption of sparsity in the current set-up, estimating these coefficients directly using numerical quadrature is problematic. Following \cite{muller2005fpca}, which proposes an estimator in the setting of sparsely observed longitudinal data, a stable estimate of $\xi_k$ can be constructed by deriving the conditional distribution of $P(\boldsymbol{\xi}_K|\boldsymbol{s}_M,\boldsymbol{P}_M)$, where $\boldsymbol{\xi}_K = (\xi_1,...,\xi_K)^\intercal$.
\par 
Given the model for the observed data \eqref{eqn:statistical_model} and Gaussian assumptions, it is easy to show that the joint distribution on $(\boldsymbol{\xi}_K, \boldsymbol{s}_M)|\boldsymbol{P}_M$ is normal with mean vector 
$
\left(\mathbf{0}_K,\boldsymbol{\mu}_M \right)^\intercal
$
and covariance matrix 
$$
\begin{bmatrix} 
\boldsymbol{\Lambda}_K &  \boldsymbol{\Lambda}_K\boldsymbol{\Psi}_{M,K}^\intercal \\
\boldsymbol{\Psi}_{M,K}\boldsymbol{\Lambda}_K & \boldsymbol{\Psi}_{M,K}\boldsymbol{\Lambda}_K\boldsymbol{\Psi}_{M,K}^\intercal + \sigma^2 \boldsymbol{I}_M\\
\end{bmatrix},$$
where $\boldsymbol{\mu}_M = (\mu(p_1), ..., \mu(p_M))^\intercal$, $\boldsymbol{\Lambda}_K = \text{Diag}(\rho_1, ..., \rho_K)$, and $\boldsymbol{\Psi}_{M,K}\in\mathbb{R}^{M\times K}$ such that $[\boldsymbol{\Psi}_{M,K}]_{mk} = \psi_k(p_m)$ for $m=1,...,M$ and $k=1,...,K$. Therefore, $P(\boldsymbol{\xi}_K|\boldsymbol{s}_M, \boldsymbol{P}_M)$ is a normal distribution with mean
\begin{equation}\label{eqn:conditional_expectation_coefficients}
    \mathbb{E}[\boldsymbol{\xi}_K|\boldsymbol{s}_M, \boldsymbol{P}_M] = \boldsymbol{\Lambda}_K \boldsymbol{\Psi}_{M,K}^\intercal [\boldsymbol{\Psi}_{M,K}\boldsymbol{\Lambda}_K\boldsymbol{\Psi}_{M,K}^\intercal + \sigma^2 \boldsymbol{I}_M]^{-1} (\boldsymbol{s}_M - \boldsymbol{\mu}_M),  
\end{equation}
where $\boldsymbol{I}_M$ is the $M\times M$ identity matrix. Denote the vector of eigenfunctions $\boldsymbol{\psi}_K(p) = (\psi_1(p), ..., \psi_K(p))^\intercal$, based on $\mathbb{E}[\boldsymbol{\xi}_K|\boldsymbol{s}_M,\boldsymbol{P}_M]$, we propose to estimate $X$ using it's conditional mean: 
\begin{equation}\label{eqn:conditional_expectation_function_approximation}
    \begin{aligned}
            \tilde{X}(p) = \mathbb{E}[X(p)|\boldsymbol{s}_M,\boldsymbol{P}_M] &= \mu(p) + \boldsymbol{\psi}_K(p)^\intercal\mathbb{E}[\boldsymbol{\xi}_K| \boldsymbol{s}_M,\boldsymbol{P}_M]  \\
            &= \boldsymbol{\phi}(p)^\intercal\left( \boldsymbol{u} + \boldsymbol{B}_K^\intercal\mathbb{E}[\boldsymbol{\xi}_K| \boldsymbol{s}_M,\boldsymbol{P}_M]\right)
    \end{aligned}
\end{equation}
where $\boldsymbol{u}\in\mathbb{R}^J$ is the vector of coefficients defining the approximation of the mean function $\mu$ over the spherical harmonic basis $\{\phi_1,...,\phi_J\}$ and $\boldsymbol{B}_K\in \mathbb{R}^{J\times K}$ is the matrix whose columns are the $\boldsymbol{b}_k$'s.
\par  
It is notable that \eqref{eqn:conditional_expectation_coefficients} is both the best estimate of the coefficients under the Gaussian assumption and the best linear predictor even when the Gaussian assumption is violated. In theory, this provides robustness to some deviations from the Gaussian assumption, a property that is verified in our simulations. Alternatively, the derivation above can be motivated through a fully Bayesian framework under the prior that $X\sim\text{GP}(\mu, C)$, or equivalently $\boldsymbol{\xi}_K \sim \mathcal{N}(0, \boldsymbol{\Lambda}_K)$, where it can be shown that the mean of the posterior predictive distribution for $X(p)$ is exactly \eqref{eqn:conditional_expectation_function_approximation}. Further consideration of these points is provided in the supplemental material.

\subsection{Optimal q-Space Sampling}\label{ssec:design_selection}
Given the estimator \eqref{eqn:conditional_expectation_function_approximation}, we formulate the diffusion sampling, the process of obtaining $\boldsymbol{s}_M$ in q-space, as an optimization problem: for a fixed total budget of $M$ samples, our goal is to select a set of directions $\boldsymbol{P}_{M}=(p_1,...,p_M)^\intercal$ such that the expected integrated squared error of the estimated diffusion signal, conditional on $\boldsymbol{P}_{M}$, is minimized. Denote the optimal locations as $\boldsymbol{P}^*_{M}$, which will be referred to as the optimal q-space design.  We first consider this problem for a single voxel and then extend it to a set of voxels in a region of interest.

\subsubsection{Optimal Diffusion Sampling for One Voxel} \label{sec:onevoxel}
At a particular voxel, we identify the optimal q-space design as the solution to the following optimization problem: 
\begin{equation}\label{eqn:optimal_design_problem}
    \boldsymbol{P}_{M}^{*} = \underset{\boldsymbol{P}_M\in\overset{M}{\underset{m=1}{\bigtimes}} \sphere}{\text{argmin}} \mathbb{E}\left( \int_{\sphere} \left(X(p) - \tilde{X}(p) \right)^2dp \right),
\end{equation}
where $\tilde{X}$ is the estimator from Equation (\ref{eqn:conditional_expectation_function_approximation}) which relies on $\boldsymbol{P}_{M}$, and the expectation is with respect to the distribution of $X$ at this voxel. 
Expanding the square and swapping the integral and the expectation operators, it is straightforward to show that \eqref{eqn:optimal_design_problem} can be equivalently formulated as the following maximization problem:
\begin{equation}\label{eqn:conditional_MISE_alternative}
   \boldsymbol{P}^*_{M} = \underset{ \boldsymbol{P}_M\in\overset{M}{\underset{m=1}{\bigtimes}} \sphere}{\text{argmax}} \int_{\sphere} \boldsymbol{c}_M(p)^\intercal \boldsymbol{\Gamma}_M^{-1}\boldsymbol{c}_M(p) dp,
\end{equation}
where, under the rank $K$ assumption on $C$, we have $\boldsymbol{c}_M(p) =  \boldsymbol{\Psi}_{M,K} \boldsymbol{\Lambda}_K\boldsymbol{\psi}_K(p)$ and 
$
\boldsymbol{\Gamma}_M =  \boldsymbol{\Psi}_{M,K}\boldsymbol{\Lambda}_K\boldsymbol{\Psi}_{M,K}^\intercal + \sigma^2 \boldsymbol{I}_M.
$
Substituting these definitions into \eqref{eqn:conditional_MISE_alternative} and integrating, we arrive at the following finite dimensional optimization problem:
\begin{equation}\label{eqn:optimization_under_low_rank_assumption}
       \boldsymbol{P}_{M}^{*} =  \underset{ \boldsymbol{P}_M\in\overset{M}{\underset{m=1}{\bigtimes}} \sphere}{\text{argmax }} \text{trace} (\boldsymbol{\Lambda}_K\boldsymbol{\Psi}_{M,K}^\intercal \boldsymbol{\Gamma}_M^{-1}\boldsymbol{\Psi}_{M,K}\boldsymbol{\Lambda}_K).
\end{equation}
\par 
In practice, it is often sufficient to constrain the candidate design points to be in a large but finite set $\mathcal{A}\subset\sphere$, e.g. a dense equispaced grid on $\sphere$ to measure diffusion-weighted signals. In this scenario, the problem \eqref{eqn:optimization_under_low_rank_assumption} can be written as 
\begin{equation}\label{eqn:global_optimal_design_problem}
    \boldsymbol{P}^{*}_M = \argmax_{\boldsymbol{P}_M\in\mathcal{P}_M(\mathcal{A})}{g(\boldsymbol{P}_M)},
\end{equation}
where $\mathcal{P}_M(\mathcal{A}) = \{(p_1, ..., p_M)^\intercal: p_i\in \mathcal{A}\}$ and 
$$
g(\boldsymbol{P}_M) = \text{trace} (\boldsymbol{\Lambda}_K\boldsymbol{\Psi}_{M,K}^\intercal \boldsymbol{\Gamma}_M^{-1}\boldsymbol{\Psi}_{M,K}\boldsymbol{\Lambda}_K).
$$ 
The global problem \eqref{eqn:global_optimal_design_problem} is an integer program and is known to be NP-hard. To circumvent this computational bottleneck, we propose a sequential search strategy for producing a greedy approximation $\widehat{\boldsymbol{P}}_{M}^{*} = (\hat{p}_1,...,\hat{p}_M)^\intercal$. Assuming that we have selected $\hat{p}^*_1, ..., \hat{p}^*_{m-1}$ in the previous rounds, we propose solving the conditional optimization problem to obtain $\hat{p}^*_m$,
\begin{equation}\label{eqn:conditional_MISE_reduced_finite}
    \hat{p}_{m}^* = \underset{p_m\in\mathcal{A}}{\text{argmax }} \text{trace} (\boldsymbol{\Lambda}_K\boldsymbol{\Psi}_{m,K}^\intercal \boldsymbol{\Gamma}_m^{-1}  \boldsymbol{\Psi}_{m,K}\boldsymbol{\Lambda}_K),
\end{equation}
where $\boldsymbol{\Psi}_{m,K}\in\mathbb{R}^{m\times K}$ is the eigenfunction evaluation matrix over $(\hat{p}_1^*, ...,\hat{p}_{m-1}^*, p_m)$ and $\boldsymbol{\Gamma}_m = \boldsymbol{\Psi}_{m,K}\boldsymbol{\Lambda}_K\boldsymbol{\Psi}_{m,K}^\intercal + \sigma^2 \boldsymbol{I}_m$. 
\par 
Algorithm \ref{alg:greedy_design_selection} presents an efficient procedure for sequentially solving \eqref{eqn:conditional_MISE_reduced_finite} by making use of a rank one update scheme to compute $\boldsymbol{\Gamma}_m^{-1}$ that leverages the block matrix inversion formula to avoid a series of costly matrix inversions, especially when $m$ gets large. {Specifically, it is easy to see that $\boldsymbol{\Gamma}_m$ can be partitioned as 
$$
\boldsymbol{\Gamma}_m = 
\begin{pmatrix}
\boldsymbol{\Gamma}_{m-1} & \boldsymbol{h} \\
\boldsymbol{h}^\intercal & q
\end{pmatrix},
$$
where $\boldsymbol{h} =\boldsymbol{{\Psi}}_{m-1, K}\boldsymbol{\Lambda}_K \boldsymbol{\psi}_K(p_{{m}}) \in\mathbb{R}^{m-1}$ and $q = \boldsymbol{\psi}_K(p_{{m}})^\intercal \boldsymbol{\Lambda}_K \boldsymbol{\psi}_K(p_{{m}}) + \sigma^2 \in \mathbb{R}$ with $\boldsymbol{\psi}_K({p_{m}}) = (\psi_1(p_{m}),...,$ $\psi_K(p_{m})) ^\intercal$, and $p_{m}$ is a candidate point in $\mathcal{A}$ for $\hat{p}^*_m$. As a result, the inverse of $\boldsymbol{\Gamma}_m$ can be computed efficiently through
\begin{equation}\label{eqn:rank_1_update}
  \boldsymbol{\Gamma}_m^{-1} = 
    \begin{pmatrix} 
    \boldsymbol{\Gamma}_{m-1}^{-1} (\boldsymbol{I}_{m-1} + a\boldsymbol{h}\boldsymbol{h}^\intercal\boldsymbol{\Gamma}_{m-1}^{-1} ) & -a\boldsymbol{\Gamma}_{m-1}^{-1} \boldsymbol{h}  \\
    -a\boldsymbol{h}^\intercal\boldsymbol{\Gamma}_{m-1}^{-1}  & a 
    \end{pmatrix},
\end{equation}
where $a = (q - \boldsymbol{h}^\intercal\boldsymbol{\Gamma}_{m-1}^{-1} \boldsymbol{h})^{-1}\in\mathbb{R}$, and  $\boldsymbol{\Gamma}_{m-1}^{-1}$ is precomputed based on the iteration of finding $\hat{p}_{m-1}$.}
\par 
A question that naturally arises is: how good is the approximation $\widehat{\boldsymbol{P}}_{M}^{*}$? In Theorem \ref{thrm:greedy_bounds}, we provide a bound for the ratio of the performance between the greedy approximation from Algorithm \ref{alg:greedy_design_selection} and and the global optimum from \eqref{eqn:global_optimal_design_problem}.
\begin{theorem}\label{thrm:greedy_bounds}
Let $\widehat{\boldsymbol{P}}_m^*$ be the approximate design obtained after $m$ iterations of Algorithm \ref{alg:greedy_design_selection}. Define $\lambda_{\psi}^{*} = \max_{p\in \mathcal{A}}\lambda_{max}(\boldsymbol{\psi}_{K}(p) \boldsymbol{\psi}_{K}(p)^\intercal)$, where $\lambda_{max}$ is the function which returns the maximum eigenvalue of its argument. Then 
\begin{equation}\label{eqn:optimal_bound_4_greedy_solution}
    g(\widehat{\boldsymbol{P}}_m^*) \ge \left[ 1 - \exp\left(\frac{-1/\rho_1}{1/\rho_K + \frac{m}{\sigma^2}\lambda_{\psi}^{*}}\frac{m}{M}\right)\right]g(\boldsymbol{P}_M^{*}), 
\end{equation}
where $\boldsymbol{P}_M^{*}$ is the solution to \eqref{eqn:global_optimal_design_problem}.
\end{theorem}
For a proof of Theorem \ref{thrm:greedy_bounds} and an interpretation of \eqref{eqn:optimal_bound_4_greedy_solution} in terms of the distribution of the random function $X$, see the supplemental material.

\begin{algorithm}
\caption{Algorithm for greedy approximation to \eqref{eqn:optimal_design_problem}}\label{alg:greedy_design_selection}
\begin{algorithmic}[1]
\State \textbf{Input}: Diagonal matrix of eigenvalues $\boldsymbol{\Lambda}_K$, eigenfunctions $\{\psi_1,...,\psi_k\}$, possible design points $\mathcal{A}$,  measurement error variance $\sigma^2$, and total budget $M\in\mathbb{N}$.
\State \textbf{Output}: Vector of indices defining the design $\boldsymbol{i}$
\State \textbf{Initialize}: $\boldsymbol{i} \gets \text{zeros}(M)$, list of indices of $\boldsymbol{a} = (1, 2, ..., |\mathcal{A}|)$
\For {$m = 1:M$}
\State $v_{optim}\gets 0$  
\For{n in $\boldsymbol{a}$}
    \State $\boldsymbol{\psi}_{n} \gets (\psi_1(p_n),...,\psi_K(p_n))^\intercal\text{ for } p_n \in \mathcal{A}$
    \If{m = 1}\Comment{handle boundary case}
    \State $\boldsymbol{\Psi}_n \gets \boldsymbol{\psi}_{n}^\intercal$; \quad $\boldsymbol{\Gamma}_n^{-1} \gets \left[ \boldsymbol{\Psi}_n \boldsymbol{\Lambda}_K\boldsymbol{\Psi}_{n} ^\intercal + \sigma^2\right]^{-1}$ 
    \Else
    \State compute $\boldsymbol{\Gamma}_n^{-1}$ using  formula \eqref{eqn:rank_1_update} and \\ \qquad\qquad\qquad let $\boldsymbol{\Psi}_n \gets ( \boldsymbol{\Psi}_{m-1,K}, \boldsymbol{\psi}_{n}^\intercal)^\intercal$ 
    \EndIf
    \State $v_{n} \gets \text{trace}(\boldsymbol{\Lambda}_K \boldsymbol{\Psi}_n^\intercal\boldsymbol{\Gamma}_n^{-1}\boldsymbol{\Psi}_n\boldsymbol{\Lambda}_K)$
    \If{$v_{n} > v_{optim}$}
    \State $i_{optim} \gets n$;\quad $v_{optim} \gets v_{n}$;\quad $\boldsymbol{\tilde{\Psi}}_{optim} \gets \boldsymbol{\tilde{\Psi}}_{n}$; \quad $\boldsymbol{\Gamma}_{optim}^{-1} \gets \boldsymbol{\Gamma}_{n}^{-1}$
    \EndIf
\EndFor
\State $\boldsymbol{i}(m) \gets i_{optim}$ and remove $\boldsymbol{i}(m)$ from $\boldsymbol{a}$
\State $\boldsymbol{\Psi}_m \gets \boldsymbol{\tilde{\Psi}}_{optim}$; \quad $\boldsymbol{\Gamma}_m^{-1}\gets \boldsymbol{\Gamma}_{optim}^{-1}$
\EndFor
\end{algorithmic}
\end{algorithm}

\subsubsection{Extension to Multiple Voxels}
The previous formulation considers the optimal design for recovery of the diffusion signal function for a single voxel. We now discuss how to extend our construction to the case when a single q-space design is needed for a collection of voxels in a region of interest $\mathcal{V}_{s}$, e.g. the whole white matter or all voxels covering certain fiber bundles of interest. Extending the objective function in \eqref{eqn:optimal_design_problem} to perform joint optimization $\forall v\in \mathcal{V}_{s}$ may be accommodated by minimizing the sum over the region 
$$
\sum_{v\in \mathcal{V}_{s}}w_v \mathbb{E}(\int_{\sphere} ({X}_v(p) - \tilde{X}_v(p))^2dp), \text{ with } \sum_{v\in \mathcal{V}_{s}}w_v = 1 \ .
$$
If no a-priori weighting is assumed, we can let $w_v=|\mathcal{V}_{s}|^{-1}$, and by similar derivations as in Section \ref{sec:onevoxel}, the sequential optimization problem in \eqref{eqn:conditional_MISE_reduced_finite} now becomes 
\begin{equation}\label{eqn:conditional_MISE_reduced_finite_region}
\begin{aligned}
    \hat{p}_m^{*} = \underset{p_m\in\mathcal{A}}{\text{argmax }}\sum_{v\in V_{s}} \text{trace} (&\boldsymbol{\Lambda}^{(v)}_K{\boldsymbol{\Psi}^{(v)}_{m,K}}^\intercal [\boldsymbol{\Psi}^{(v)}_{m,K}\boldsymbol{\Lambda}^{(v)}_K{\boldsymbol{\Psi}^{(v)}_{m,K}}^\intercal \\
    &+ \sigma^2 \boldsymbol{I}_m]^{-1}\boldsymbol{\Psi}^{(v)}_{m,K}\boldsymbol{\Lambda}^{(v)}_K),
\end{aligned}
\end{equation}
where the superscript $v$ indicates the voxel specific data. Simple augmentation of Algorithm \ref{alg:greedy_design_selection} can be made to accommodate this situation.

\subsection{Prior Summarization From Historical Data}\label{ssec:prior_summarization}

Notice that both the estimator \eqref{eqn:conditional_expectation_function_approximation} and Algorithm \ref{alg:greedy_design_selection} depend on unknown parameters $\sigma^2$, $\boldsymbol{u}$, $\boldsymbol{B}_K$, and $\boldsymbol{\Lambda}_K$, the latter two of which can be computed from the decomposition of $\boldsymbol{\Sigma}$ (defined in equation (\ref{eqn:sigma})) along with any standard technique to select the rank $K$, e.g. the proportion of variance explained. $\sigma$ characterizes the signal to noise ratio and is related to the scanner used to acquire the data. $\boldsymbol{u}$ and  $\boldsymbol{\Sigma}$ characterize the mean and variance of the diffusion signal in a voxel. 

We propose to estimate  $\sigma$ from  $b = 0$ (non-diffusion weighted) data with the assumption that we have access to $n$ ($> 2$) such images for the subject to be scanned, e.g. collected at some pre-screening step (see the supplemental material for further elaboration and evaluation of this estimator). In the rest of this section, we discuss how to leverage  high-quality historical diffusion data to estimate ${\boldsymbol{u}}$ and $ {\boldsymbol{\Sigma}}$.

\par 
We assume that we have access to a relevant collection of densely sampled high-resolution historical dMRI data from the population of interest, denoted as $\mathcal{H} = \{(p_{im}, s_{vim}): \text{for } i = 1, ..., N, m = 1, ..., M_i, \text{ and } v = 1,..., |\mathcal{V}_i|\}$, where $p_{im}$ is a q-space acquisition direction, $M_i$ is the number of samples, $s_{vim}$ is the observed diffusion signal for the $i$th subject in the $m$th direction at the $v$th voxel and $N$ represents the number of subjects. Such data are ubiquitous and easy to access due to recent efforts from the HCP and the UK Biobank. We leverage $\mathcal{H}$ to estimate ${\boldsymbol{u}}$ and $ {\boldsymbol{\Sigma}}$ across the brain. This proves challenging in that each subject's data in $\mathcal{H}$ is gathered in a different coordinate system and the subjects have different brain morphology. We circumvent this issue by registering the historical data to a common space (also referred to as the template space). The mean and covariance functions are estimated at each voxel in the template space. The warping function used to perform registration is constrained to be a diffeomorphism. As a result, for any subject of interest that has been registered to the template space, the inverse warping can be used to map the estimated priors in template space to the subject space smoothly. We now discuss this pipeline in detail.

\subsubsection{Diffusion Signal Estimation in Subject Space}\label{sssec:diffusion_signal_estimation}
Given the fact that data in $\mathcal{H}$ are acquired with dense samples, we can confidently estimate the smoothed diffusion signal at each voxel independently using the popular regularized least squares estimator \cite{descoteaux_2007}. In brief, the diffusion signal is represented using a basis expansion over the $\phi_j$'s. For voxel $v$ in historical subject $i$, its diffusion signal is estimated by solving the following regularized least squares problem
$$
    \hat{\boldsymbol{c}}_{vi} = \underset{\boldsymbol{c}\in\mathbb{R}^J}{\text{argmin}} \overset{M_i}{\underset{m=1}{\sum}} (s_{vim} - \boldsymbol{c}^\intercal \boldsymbol{\phi}(p_{im}))^2 + \lambda_{i} \int_{\sphere} (\Delta_{\sphere}(\boldsymbol{c}^\intercal \boldsymbol{\phi}(p)))^2 dp,
$$
where $\boldsymbol{\phi}(p)=(\phi_1(p),...,\phi_J(p))^\intercal$, $\Delta_{\sphere}$ denotes the Laplace-Beltrami operator on $\sphere$ and  penalty $\lambda_{i} > 0$ is selected by minimizing the generalized cross validation (GCV) criteria. The resulting smoothed estimate of diffusion signal is given by $\widehat{X}_{vi}(p) = \hat{\boldsymbol{c}}_{vi}^\intercal \boldsymbol{\phi}(p)$. More details on this estimation procedure can be found in the supplementary materials. 

\subsubsection{Mean and Covariance in Template Space}\label{sssec:mean_cov_template}
Let $\gamma_i$ be the diffeomorphic warping function from the historical subject $i$ to the desired template space, e.g., based on the subject's T1 or fractional anisotropy image \cite{avants2008}. The warping function deforms the regular voxel grid in subject space to align the subject's data to the template,  and a re-gridding procedure is needed in the template space to reconstruct a regular voxel grid. We can obtain an estimate of the subject's diffusion function at each of the voxels in the template space through interpolation of $\widehat{X}_{vi}$. That is, for any voxel $\tilde{v}$ in the template space, we have $\widehat{X}_{\tilde{v}i}(p) = \hat{\boldsymbol{c}}_{\tilde{v}i}^\intercal\boldsymbol{\phi}(p)$ where $\hat{\boldsymbol{c}}_{\tilde{v}i}$ is computed by a linear interpolation scheme applied to $\{\hat{\boldsymbol{c}}_{vi}\}$ for $v$'s in the subject space mapped near $\tilde{v}$ by $\gamma_i$. We use the popular Advanced Normalization Tools (ANTs) to perform this alignment and refer the reader to \cite{avants2009advanced} for more details on the interpolation procedure.
\par  
Once the aligned diffusion signal $\widehat{X}_{\tilde{v}i}(p)$ has been computed for all subjects in $\mathcal{H}$, the mean and covariance functions at $\tilde{v}$ in the template space can be estimated by the empirical estimators: $\hat{\mu}_{\tilde{v}}(p) =  {\hat{\boldsymbol{u}}}_{\tilde{v}}^\intercal \boldsymbol{\phi}(p)$ and $\widehat{C}_{\tilde{v}}(p, t) = \boldsymbol{\phi}(p)^\intercal \widehat{\boldsymbol{\Sigma}}_{\tilde{v}} \boldsymbol{\phi}(t)$, where ${\hat{\boldsymbol{u}}}_{\tilde{v}}$ and $\widehat{\boldsymbol{\Sigma}}_{\tilde{v}}$ are the sample mean and covariance matrix of $\{\hat{\boldsymbol{c}}_{\tilde{v}1}, ..., \hat{\boldsymbol{c}}_{\tilde{v}N}\}$, respectively.
We assume a large historical sample, $N > J$ and therefore we expect these estimators to be stable. 

\subsubsection{Mean and Covariance in Subject Space}\label{sssec:mean_cov_subject}
\par 
Let $\gamma$ be the warping function to align the new subject to be scanned to the template space. GP($\hat{\mu}_{\tilde{v}}$, $\widehat{C}_{\tilde{v}}$) characterizes the distribution of the diffusion signal at voxel $\tilde{v}$ in the template space. Our goal is to map the GPs in the template space to  the subject space to facilitate q-space sampling (Algorithm \ref{alg:greedy_design_selection}) and diffusion signal estimation (estimator \eqref{eqn:conditional_expectation_function_approximation}). To do this, we estimate the mean and covariance functions at each voxel in subject space using linear interpolation guided by the inverse warping transformation $\gamma^{-1}$. This process is described in detail below.
\par 
The mean functions are parameterized by ${\hat{\boldsymbol{u}}}_{\tilde{v}}$, a vector in $\mathbb{R}^J$, and therefore ${\hat{\boldsymbol{u}}}_{v}$ for subject space voxel $v$ can be formed by standard multivariate linear interpolation. The covariance functions are parameterized by symmetric positive definite (SPD) matrices $\widehat{\boldsymbol{\Sigma}}_{\tilde{v}}$ and thus the estimate of $\widehat{\boldsymbol{\Sigma}}_v$ for subject space voxel $v$ requires an interpolation of SPD matrices. Considering the non-Euclidean structure of the SPD matrix manifold, we propose the following interpolation scheme.
\par 
Denote $\mathcal{S}_{+}^J$ the manifold of $J\times J$ SPD matrices and let $\boldsymbol{M}_1, ..., \boldsymbol{M}_n$ with each $\boldsymbol{M}_l\in\mathcal{S}_{+}^J$. The weighted Karcher mean of the sample is defined as 
\begin{equation}\label{eqn:karcher_mean}
        \boldsymbol{M}_{Kmean} = \argmin_{\boldsymbol{M}\in\mathcal{S}_{+}^J}\sum_{i=1}^n w_id^2(\boldsymbol{M}_i, \boldsymbol{M}), \text{ with  }\sum_{i=1}^nw_i = 1,
\end{equation}
where $d(\cdot,\cdot)$ is induced by a selected Riemannian metric. 
If we let the weights $w_i$ be proportional to the Euclidean distance between $\gamma^{-1}(\tilde{v}_i)$ and the voxel of interest $v$ in the subject space, the weighted Karcher mean can be used as an interpolation scheme.
\par 
There are several valid metrics that can be used to compute the distance between elements in $\mathcal{S}_{+}^J$ \cite{arsingy2007}. It is easy to show $\int_{\mathbb{S}^2\times \mathbb{S}^2} (\widehat{C}_{\tilde{v}_{1}}(p, t) - \widehat{C}_{\tilde{v}_{2}}(p, t))^2dpdt = \|\widehat{\boldsymbol{\Sigma}}_{\tilde{v}_{1}} - \widehat{\boldsymbol{\Sigma}}_{\tilde{v}_{2}} \|_{F}^2$, so using the standard Euclidean (Frobenius) norm as a metric for the interpolation appears as a reasonable choice. However, interpolation of SPD matrices under the Euclidean norm can result in a ``swelling'' effect, in which the determinant of the interpolated value is larger than the determinant of any of the original SPD matrices \cite{dryden2009}. This would result in an undesirable amplification of the estimated variability. To overcome this limitation, we use the log-Euclidean metric proposed in \cite{arsingy2007}
$$
    \|\boldsymbol{M}_1 - \boldsymbol{M}_2\|_{Log}^2 = \|\text{log}(\boldsymbol{M}_1) - \text{log}(\boldsymbol{M}_2) \|_{F},
$$
where $\text{log}(\cdot)$ represents the matrix logarithm. For any $\boldsymbol{M}\in\mathcal{S}_{+}^J$, the matrix logarithm always exists and is defined by $\text{log}(\boldsymbol{M}) = \boldsymbol{V} \text{log}(\boldsymbol{D}) \boldsymbol{V}^\intercal$, where $\boldsymbol{V}$ is the matrix whose columns are the eigenvectors of $\boldsymbol{M}$ and $\boldsymbol{D}$ is the diagonal matrix of eigenvalues. Since $\boldsymbol{D}$ is diagonal, the matrix logarithm has an especially simple form: $\text{log}(\boldsymbol{D}) = \text{Diag}(\text{log}(\boldsymbol{D}_{11}), \text{log}(\boldsymbol{D}_{22}), ..., \text{log}(\boldsymbol{D}_{JJ}))$. The subject space covariance function for voxel $v$ is defined by the solution to the interpolation \eqref{eqn:karcher_mean} applied to the set of $\widehat{\boldsymbol{\Sigma}}_{\tilde{v}}$, for $\tilde{v}$ mapped near $v$ by $\gamma^{-1}$, with $d(\boldsymbol{M}_1,\boldsymbol{M}_2) = \|\boldsymbol{M}_1 - \boldsymbol{M}_2\|_{Log}^2$. 

With the estimated $\sigma$, ${\boldsymbol{u}}$ and ${\boldsymbol{\Sigma}}$ for the subject to be scanned, we now can apply  Algorithm \ref{alg:greedy_design_selection} to acquire sparsly sampled dMRI data, and estimator \eqref{eqn:conditional_expectation_function_approximation} to recover the diffusion signal (and fODF with the FRT). 

\section{Simulation Study}\label{sec:simulation}

In this study, we compare the performance of the proposed method with the penalized linear regression approach using spherical harmonics from \cite{descoteaux_2007} (denoted SHLS) and the spherical finite rate of innovation and sparse regularization estimator from \cite{deslauriers2013} (denoted SFR) over a series of increasing budgets. For a given budget, the samples for SHLS and SFR were selected using the popular electrostatic repulsion model (ESR). The ESR algorithm constructs a sampling design by modeling the acquisition directions as points on $\sphere$ and attempting to find a configuration which minimizes the electrostatic energy between point pairs. It has been widely used  for constructing acquisition schemes over $\sphere$, e.g., in the HCP.  The proposed greedy approach (Algorithm \ref{alg:greedy_design_selection}), denoted GDS (greedy design selector), was used to select the sampling design for our sparse estimator \eqref{eqn:conditional_expectation_function_approximation}, denoted CU (conditional expectation update).
\par
\begin{figure}
    \centering
    \includegraphics[scale=0.58]{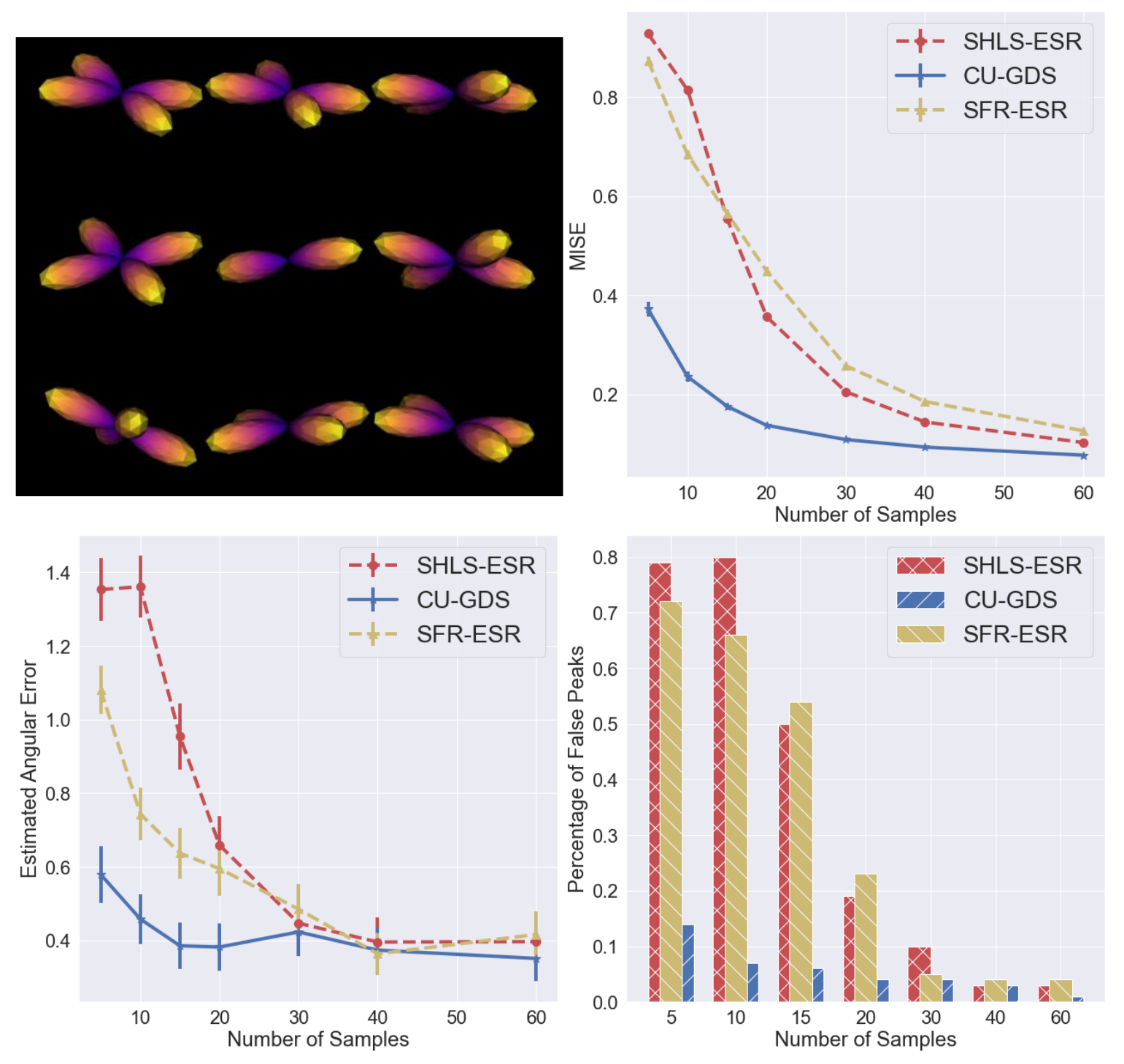}
    \caption{Simulation study results. The top left panel displays a sample of fODFs generated from model~\eqref{eqn:sim_gen_model}. The top right figure compares the MISE of different methods. The bottom two panels compare the angular accuracy, where the bottom left panel displays the percentage of false peaks and the bottom right gives the angular error for a sequence of increasing budgets.}
    \label{fig:simulation_figure}
\end{figure}

 
We simulate the diffusion signal at a particular voxel with both single and crossing fibers as follows. Denote $g_{\boldsymbol{m},\kappa}$ the density function of the von Mises-Fisher (VMF) distribution with mean direction $\boldsymbol{m}\in\mathbb{S}^2$ and concentration parameter $\kappa$. A sample fODF is simulated according to the generative model 
\begin{equation}\label{eqn:sim_gen_model}
\begin{aligned}
        f_i &\quad \propto \quad  w_1(g_{\boldsymbol{m}_{1i},10} + g_{-\boldsymbol{m}_{1i},10}) + w_2(g_{\boldsymbol{m}_{2i},10} + g_{-\boldsymbol{m}_{2i},10})\\
        \boldsymbol{m}_{1i} &\sim \text{VMF}\left(\boldsymbol{\nu}_1, 20\right) \quad \boldsymbol{m}_{2i} \sim \text{VMF}\left(\boldsymbol{\nu}_2, 20\right)
\end{aligned}
\end{equation}
where $\boldsymbol{\nu}_1=(1,0,0)$, $\boldsymbol{\nu}_2=(\frac{1}{\sqrt{3}},-\frac{1}{6}(3-\sqrt{3}),\frac{1}{6}(3+\sqrt{3}))$, and $w_1 = w_2 = 0.5$. Note that we must include $g_{-\boldsymbol{m}_{ji,10}}$ to respect the symmetry of the diffusion process. Some examples of generated fODFs from \eqref{eqn:sim_gen_model} are shown in the top left panel of Figure~\ref{fig:simulation_figure}. Note that single fibers are simulated from \eqref{eqn:sim_gen_model} when the samples $ \boldsymbol{m}_{1i}$ and $ \boldsymbol{m}_{2i}$ are sufficiently close. The sampled fODF is then represented using the real, symmetric spherical harmonics $\{\phi_1, ...,\phi_{J}\}$. Under this representation, the FRT is invertible (i.e. it is a bijection for symmetric differentiable functions on $\mathbb{S}^2$) and has a simple closed form which can be applied to the coefficients of $\{\phi_1, ...,\phi_{J}\}$ to obtain the corresponding diffusion signal $X_i$.
\par 
$N=200$ diffusion signals were simulated according to this procedure and evaluated at 90 locations on $\mathbb{S}^2$ selected by the ESR algorithm. Each observation was contaminated with i.i.d. measurement error with $\sigma = 0.01$. An independent test set of size $N_{test} = 100$ was also constructed using the same method. The reconstruction techniques were applied to the test set for a sequence of increasing observations (budgets). SFR estimates the fODFs directly, so an estimate of the diffusion signal function was obtained by applying the inverse FRT. In contrast, SHLS and CU estimate the diffusion signal function and therefore we apply the FRT and then the spherical deconvolution \cite{descoteaux_2009} to obtain the fODF estimates.
\par 
Performance was compared using three evaluation metrics: the mean integrated squared error (MISE) of the estimated diffusion signal, the percentage of false peaks (PFP), and the estimated angular error (EA). The PFP and EA are popular measures quantifying the angular fidelity of the estimated fODF. The PFP is computed simply as the proportion of estimated fODFs in the test set which have the same number of peaks as the true fODF, i.e.
$$
    \text{PFP} = N_{test}^{-1}\sum_{i=1}^{N_{test}} \mathbb{I}\{\text{np}(\hat{f}_i) = \text{np}(f_i)\}
$$
where np is a function that counts the number of local maximum of a $\mathbb{L}^2(\mathbb{S}^2)$ function. We defined the estimated angular error (EA) as 
$$
    \text{EA} = N_{test}^{-1}\sum_{i=1}^{N_{test}} |\theta(\hat{f}_i) - \theta(f_i)|
$$
where $\theta$ is the function defined as follows: if only a single peak is detected, $\theta = 0$; if two or more peaks are detected, $\theta$ is set to the angle between the two highest peaks. This is a natural extension of the EA metric used in \cite{ning2015}. 
\par 
The top right panel of Figure~\ref{fig:simulation_figure} shows the results of the MISE comparison. The proposed method is uniformly outperforming both of the competitors over all budgets considered. The boost in performance is especially pronounced in the sparse ($\le 20$) sample case. The bottom left panel of Figure~\ref{fig:simulation_figure} displays the PFP and the bottom right shows the EA for all estimators as a function of budget. The same pattern in the MISE evaluation is observed in both the angular assessments, indicating that the boost performance for the sparse sample diffusion signal estimation carries through to quality fODF reconstruction. Additional simulation studies investigating the performance in Non-Gaussian and low signal to noise ratio scenarios result in similar conclusions to those presented. Further, we
found the computational complexity of CU-GDS fitting to
be comparable with that of SHLS-ESR. For these results as well as a discussion on the selection of an adequate training sample size $N$, we refer the interested reader to the supplemental material.

\section{Real Data Experiments}\label{sec:real_data}

\subsection{Data Acquisition and Preprocessing}

We evaluated our method using diffusion data from two datasets: HCP and a local aging dataset consisting of older adults with mild cognitive impairment (MCI). The full HCP imaging acquisition and minimal diffusion MRI image preprocessing steps for HCP data are documented in \cite{glasser_2013}. Briefly, all imaging was conducted on the 3T Siemens Connectom scanner (Erlangen, Germany). High-resolution T1w anatomical images were acquired with the 3D MPRAGE (magnetization prepared rapid gradient echo) sequence with a slice acceleration factor of 2 using 0.7 mm isotropic resolution. Diffusion imaging was performed using a 2D spin-echo EPI (echo planar imaging) sequence with approximately 90 diffusion directions at three non-zero b-values (1,000, 2,000, and 3,000 s/mm$^{2}$ each) and 18 $b=0$ reference scans at 1.25 mm isotropic resolution. A full diffusion MRI run includes 6 runs of about 9 mins 50 seconds each, representing 3 acquisitions of different b-vectors, each acquired once with right-to-left (RL) and left-to-right (LR) phase encoding polarities; these are used to correct for susceptibility-induced distortion.
\par
For our study, we randomly selected 240 healthy HCP subjects (age range: 20-35 years) to be used as training data, i.e. high-resolution historical data $\mathcal{H}$. An additional 10 subjects from the same demographic were used as independent test data. A template of the diffusion data was created from the FA images of 20 randomly selected training subjects. Each subject's FA image was then warped to the template using the nonlinear registration algorithm in ANTs \cite{avants2009advanced}. The training data were used to estimate 
two template space priors according to the procedure outlined in Section \ref{ssec:prior_summarization} using the $b=1,000$ and $b=2,000$ shells, respectively. The inverse warping function for each subject in the test data was used to map the template space mean and covariance functions into their subject space.
\par 
To study the generalizability of our method, we utilized a local dataset consisting of 16 older adults ($ > 60$ years of age) with MCI. The T1 and dMRI data were collected using a Siemens 3T Prisma scanner (Erlangen, Germany) at the University of Rochester Imaging Center. A total of 14 $b=0$, 26 $b=1,000$ and 102 $b=2,000$ images were obtained with q-space locations selected using the ESR algorithm. All acquisitions were acquired with the same $2$ mm isotropic resolution. Susceptibility distortion and motion correction were performed using FSL. The T1 image was processed using ANTs and FreeSurfer, registered to the dMRI and a white matter mask was extracted. For additional details about the study population and scanning protocol, see \cite{Venkataraman2021}.

\subsection{Global Estimation and Tractography}
All experiments in this subsection are based on the HCP data. 
For a sequence of increasing total budgets, we applied the GDS algorithm to select diffusion directions from the set of 90 $b=1,000$ vectors for each HCP test subject independently. We determined an optimal set of directions to sample 1) for {\it each voxel}, and 2) for {\it an ROI}. For the second case, we randomly selected $10,000$ voxels within the white matter mask to define the ROI $\mathcal{V}_{s}$, i.e. approximating the optimal directions over the whole white matter. In analysis not reported here, we observed that the final results were insensitive to resamplings of $\mathcal{V}_{s}$, so long as the sample was sufficiently large. For each design selected by GDS, we estimated the diffusion signal using the proposed estimator \eqref{eqn:conditional_expectation_function_approximation}, denoted again as CU. Throughout this section, we compare our method to SHLS-ESR for various tasks of interest, noting that it displayed largely similar performance to SFR-ESR in the simulation study and was found to be much faster for large datasets.
\par
\begin{figure}
    \centering
    \setlength{\tabcolsep}{1pt}
    \begin{tabular}{cc}
    \includegraphics[scale=0.55]{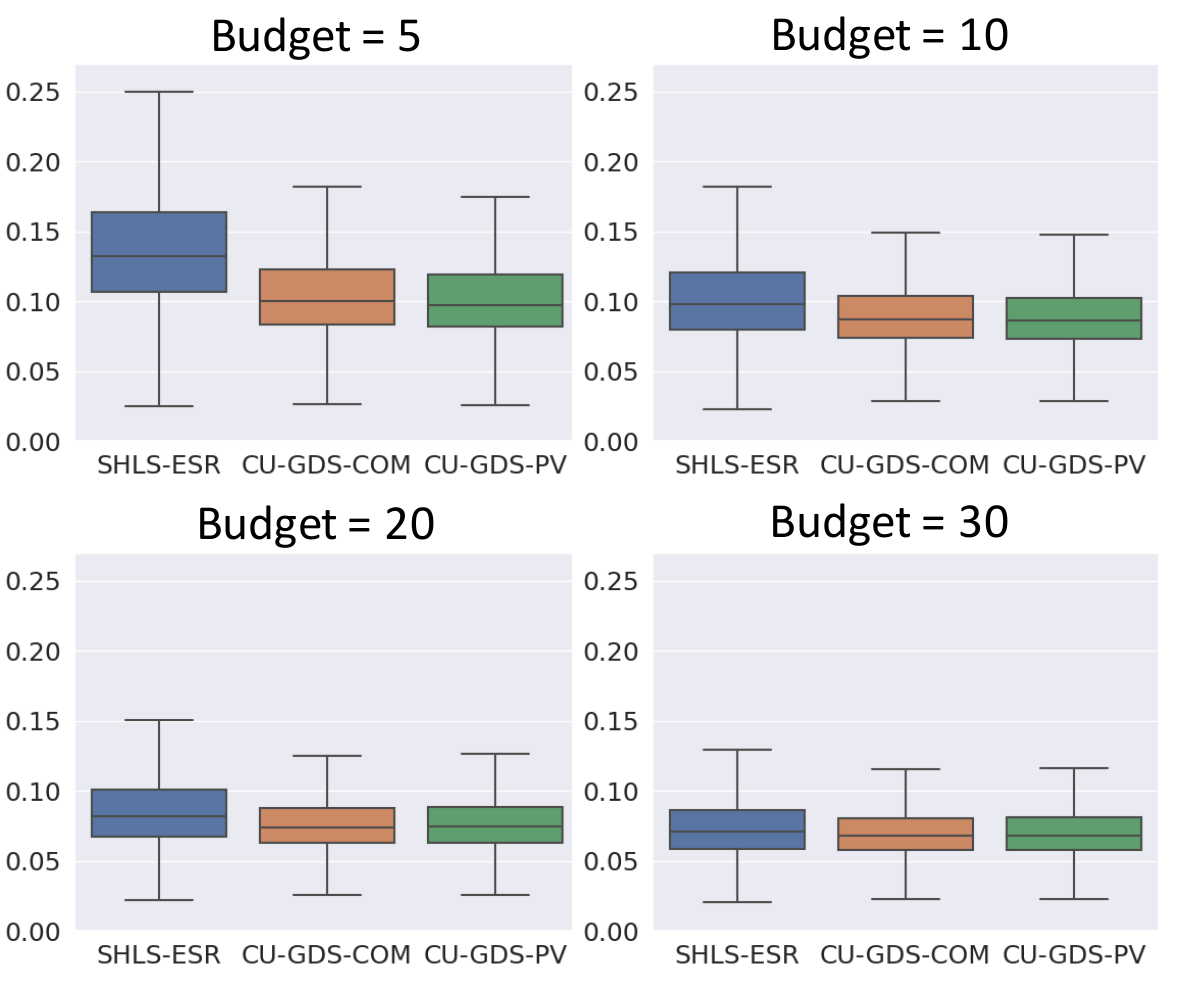} \\
    (a) \\
    \includegraphics[scale=0.4]{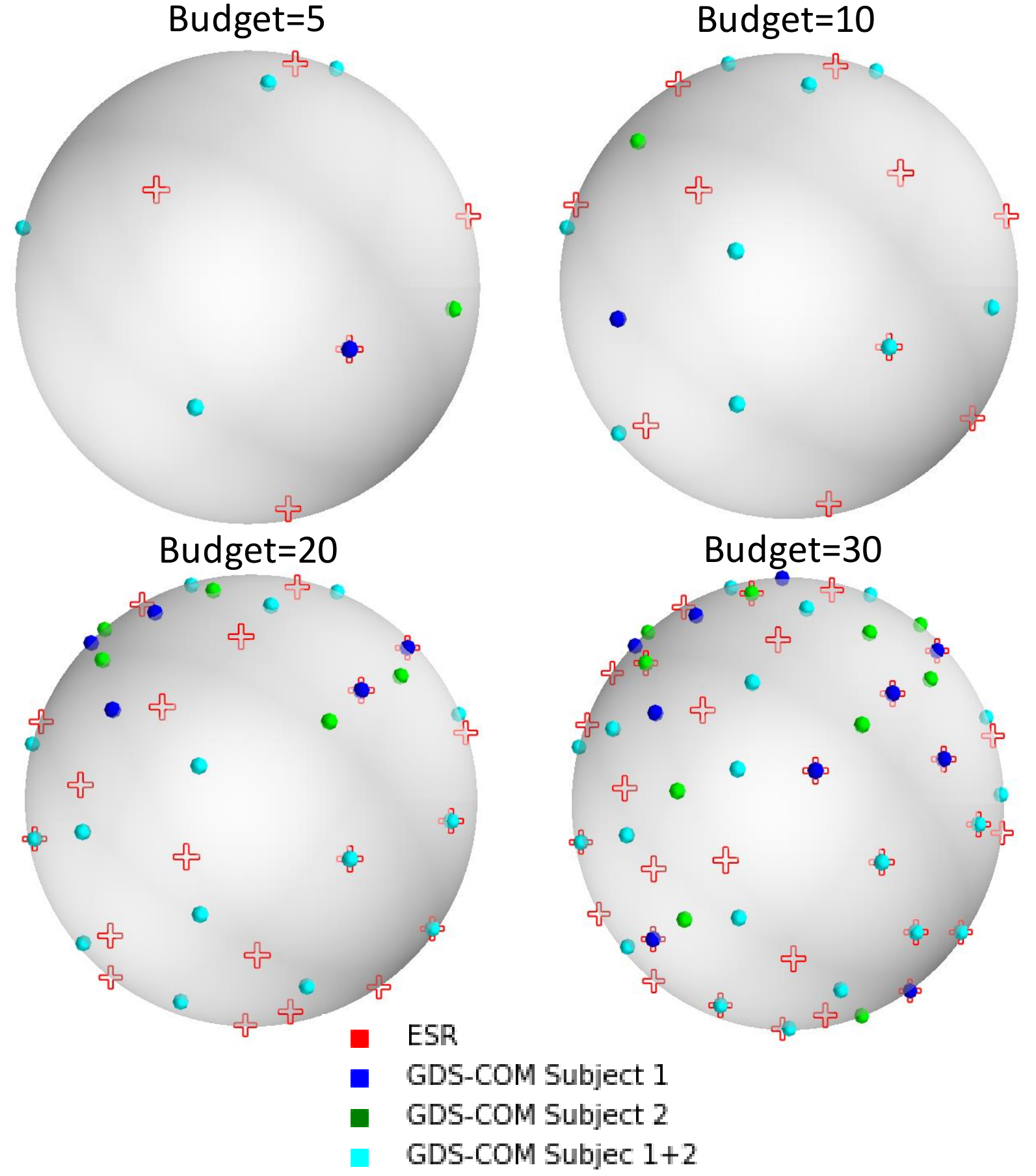} \\
    (b)\\ 
    \end{tabular}
    \caption{(a) Boxplots of the MISE between the sparse sample reconstructions (SHLS-ESR, CU-GDS-COM, and CU-GDS-PV) and SHLS-ESR90, where CU-GDS-COM represents CU fit based on per subject design, and CU-GDS-PV represents CU fit based on per voxel design. (b) Comparison of two q-space sampling schemes, ESR and our GDS-COM, for two randomly selected test subjects from HCP. The sample location for ESR are denoted by red crosses. The cyan points indicate sample locations selected by the GDS-COM for both of the test subjects, while the blue and green samples are locations selected for only one of the subjects. Each panel shows the selected design under a different budget constraint.}
    \label{fig:signal_error_distribution}
\end{figure}

 
Since in general SHLS-ESR reconstruction enjoys good performance for dense samples, we first evaluated the diffusion signal estimation for both techniques at various sparse budgets by comparing them to the SHLS-ESR applied to the full 90 directions, denoted as SHLS-ESR90. Figure \ref{fig:signal_error_distribution} (a) displays boxplots of the MISE between the sparse sample reconstructions and SHLS-ESR90 over all of the  white matter voxels, where CU-GDS-PV (per voxel) represents sampling scenario 1), and CU-GDS-COM (common, per subject) represents sampling scenario 2). The CU-GDS framework produced reconstructions which better approximate the signal estimates of SHLS-ESR90 than does SHLS-ESR for each of the sparse budgets considered. The performance boost becomes more pronounced in the sparser cases, e.g., when we only have the budget to sample fewer than 20 directions. The per voxel design resulted in lower average MISE than the per subject design, as expected since the latter is equivalent to the former with an additional constraint, but the difference is minimal relative to the performance of SHLS-ESR. Figure \ref{fig:signal_error_distribution} (b) displays the optimal per-budget q-space sampling schemes constructed from the GDS-COM algorithm for two randomly selected HCP test subjects. The corresponding ESR schemes are also included for comparison (red crosses). We note two important features. First, for a given budget, the two GDS-COM designs select relatively few q-space samples in common with ESR. This indicates that, under the CU estimator, the ESR method produces sub-optimal designs for 
diffusion signal estimation. Second, the two GDS-COM designs have a relatively large number of design points in common (the cyan color). This is not surprising, as there exist large similarities in the diffusion profiles of healthy young adult brains. That said, the designs are not identical, indicating there is some residual subject-specific variability driving the selection of the locations that differ.
\par 
\par
\begin{figure}
\includegraphics[scale=0.47]{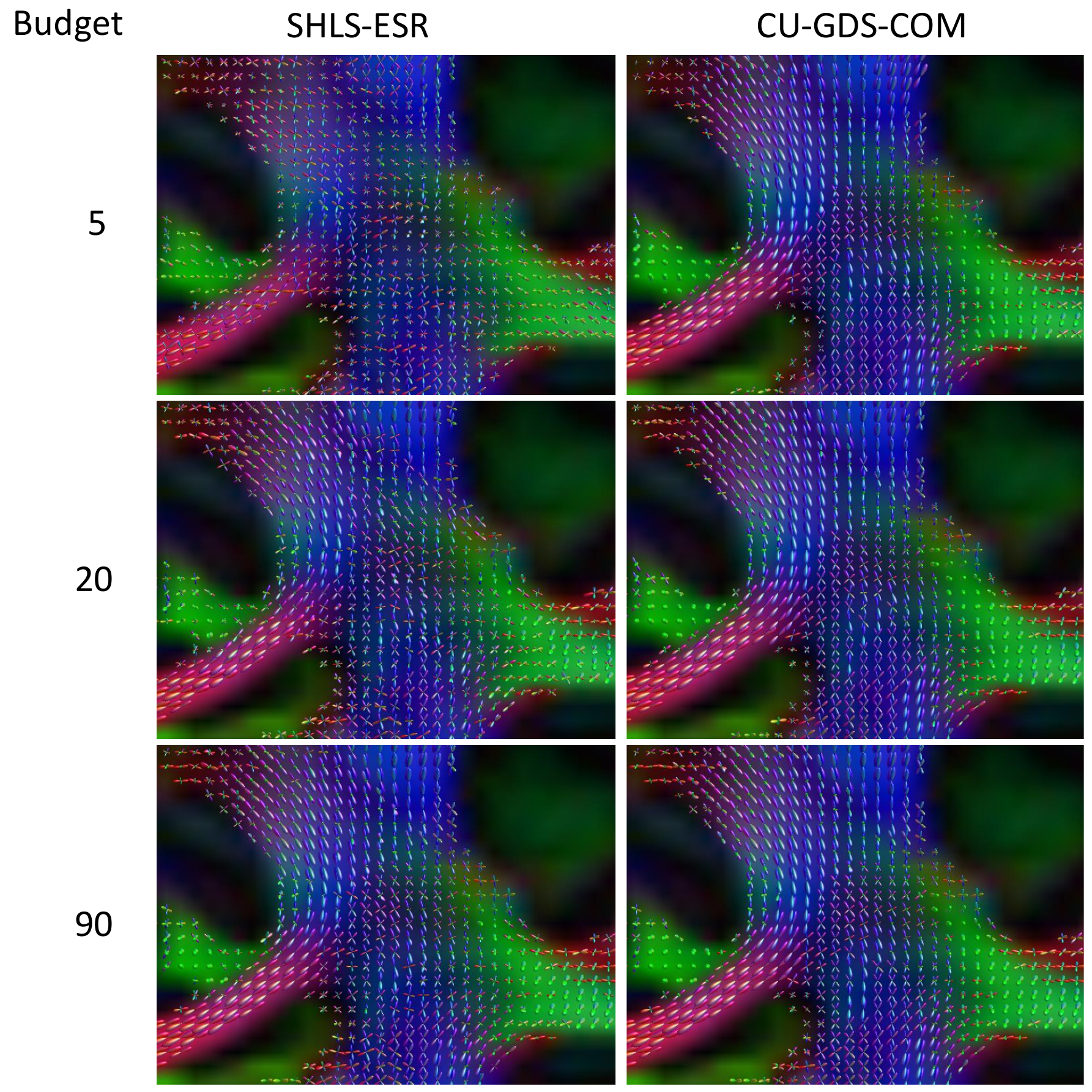}
\includegraphics[scale=0.47]{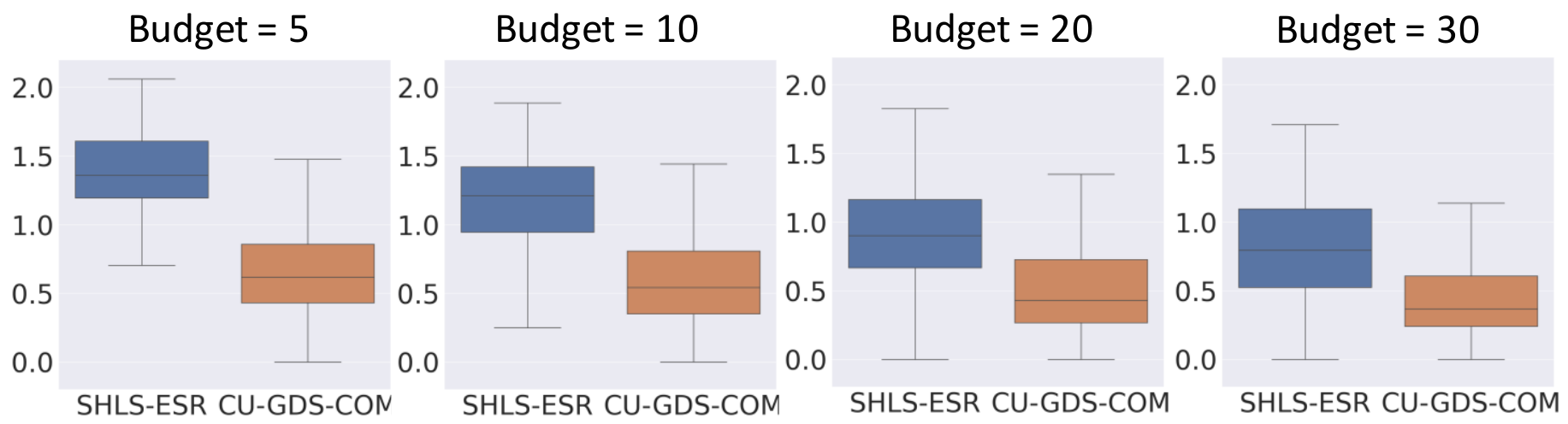}
\caption{fODFs from a coronal slice of a randomly selected test subject reconstructed by SHLS-ESR (left column) and CU-GDS-COM (right column) for several budgets considered. The boxplots display the per voxel MISE between the sparse budgets and full data reconstructions, for both methods.}
\label{fig:coronal_odf_slice}
\end{figure}

 
For each total budget considered, fODFs were constructed from the diffusion signal estimates by applying first the FRT and subsequently the spherical deconvolution \cite{descoteaux_2009}. Figure \ref{fig:coronal_odf_slice} displays the fODFs from a coronal slice of a randomly selected test subject. Using only 5 directions, CU-GDS-COM estimation already results in fODFs that very closely approximate those based on the full set of 90 directions. For instance, the single fiber region in the corpus callosum (bottom left pink area) as well as the crossing fibers in the centrum semiovale (center blue area) are nearly identical between the budget $=5$ and budget $=90$ plots. This is not the case for SHLS-ESR, which struggles in both areas for sparse budgets. Boxplots of the per-voxel MISE between the sparse budget and full data reconstructions for both methods are displayed below. For each sparse budget considered, the CU-GDS-COM produces estimates which more faithfully represent the full data reconstructions than does the SHLS-ESR. 

\begin{figure}
\setlength{\tabcolsep}{1pt}
    %\begin{tabular}{ccc}
    %    \includegraphics[scale=0.3]{Figures/numbers.pdf} & 
    %     \includegraphics[scale=0.3]{Figures/tractography_esr.pdf} & \includegraphics[scale=0.3]{Figures/tractography_gds.pdf} \\
    %\end{tabular}
    %\includegraphics[scale=0.2]{Figures/Connectome_Results.pdf}
    \begin{tabular}{c}
            \includegraphics[scale=0.35]{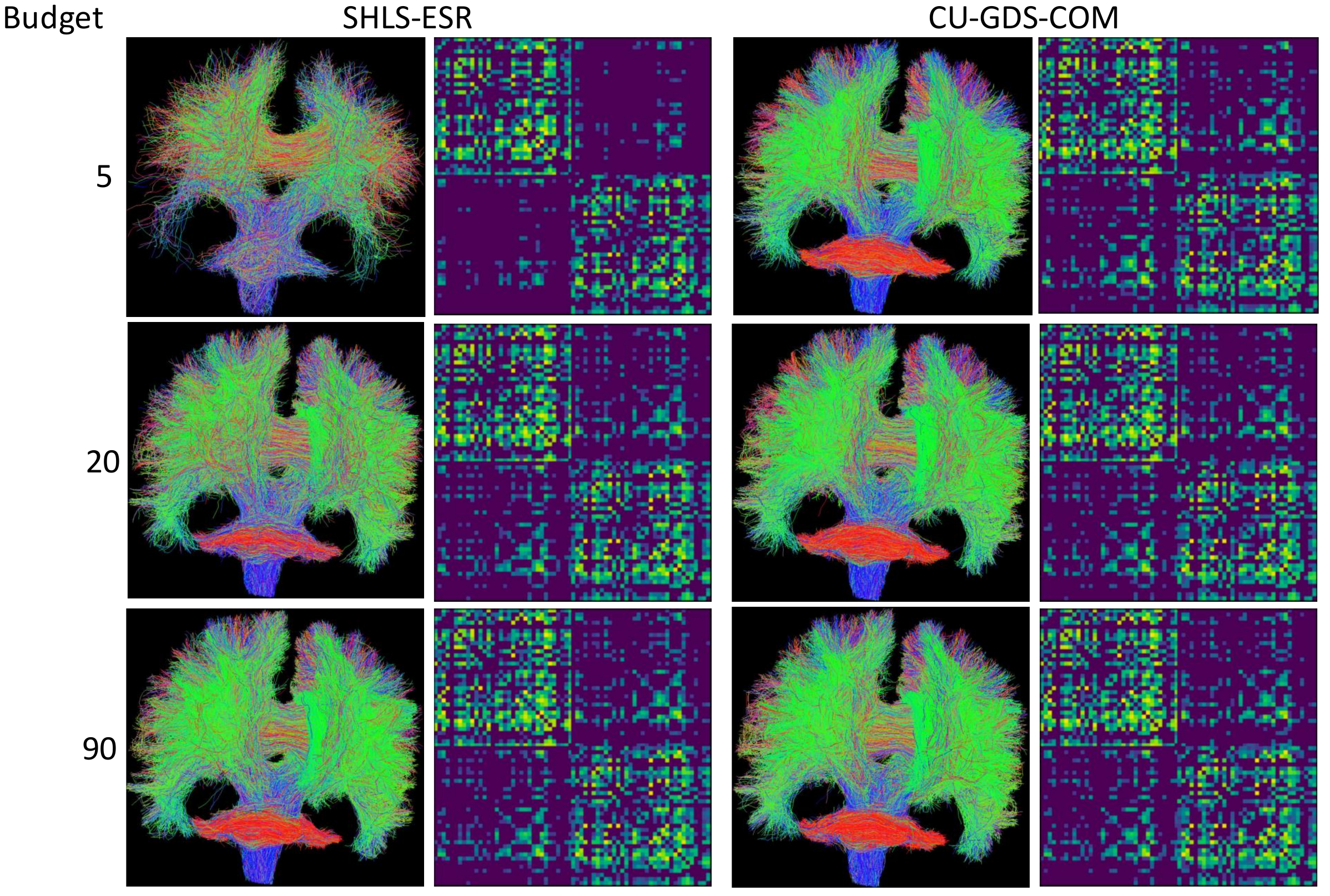}
            \\
            \includegraphics[scale=0.25]{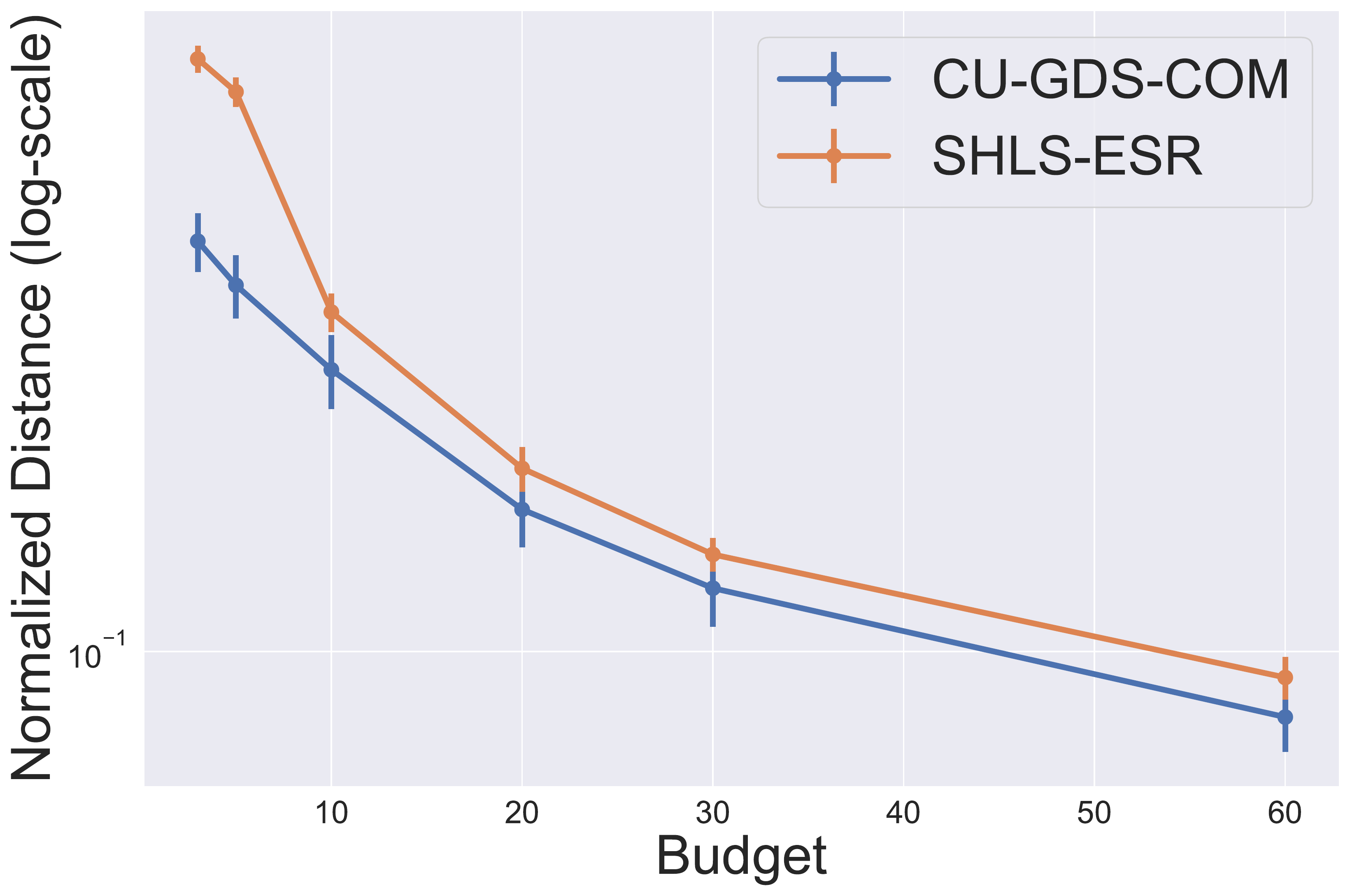} \\ 
            (a) \\
         \includegraphics[scale=0.2]{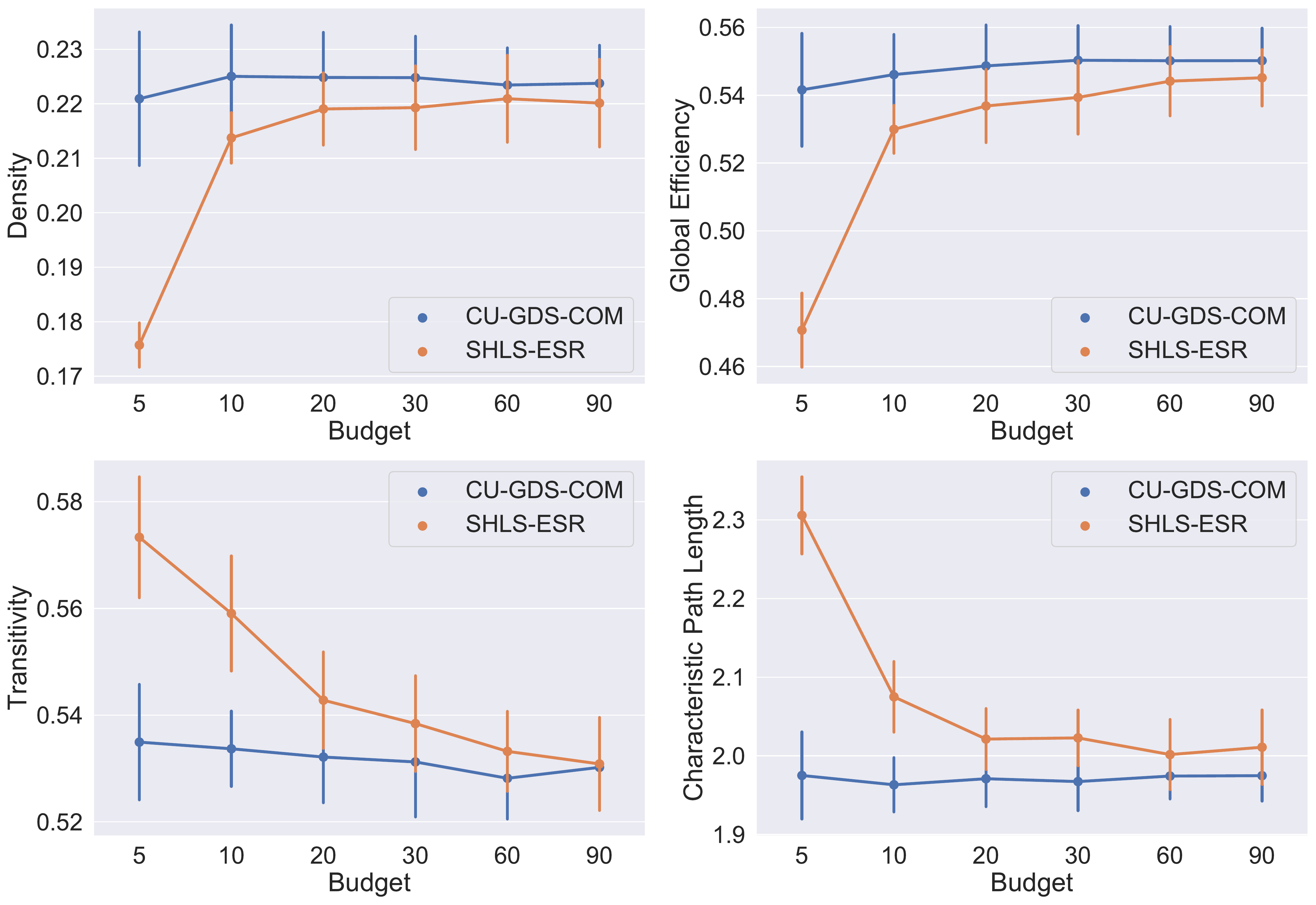} \\
         (b)
    \end{tabular}
    \caption{(a) WM fiber streamlines and connectivity matrices from a randomly selected test subject. Tractography was run on fODFs constructed from diffusion signal function estimates from SHLS-ESR (left column) and CU-GDS-COM (right column) for a sequence of increasing total budgets. The line plot shows the normalized Frobenius distance between the budget limited reconstruction and full data connectome, on the log scale. (b) Line plots of several graph theoretic metrics of brain networks for SHLS-ESR and CU-GDS-COM methods over different budgets computed on the test set. Means and $95\%$ error bars are displayed.}
    \label{fig:tractography}
\end{figure}

\par
The motivating application for this paper is to develop a sparse diffusion weighted signal sampling framework for brain structural connectome analysis. Therefore, we investigated the quality of the tractogram and structural connectivity resulting from the reconstructed fODFs at various sampling budgets. The tractogram was computed from the fODFs using the particle filtering tractography algorithm \cite{Girard2014g} that is implemented in \textit{dipy}. Connectome matrices were reconstructed using the Desikan atlas \cite{desikan_2006} created by FreeSurfer \cite{fischl_2012}. The top plot in Figure~\ref{fig:tractography} (a) shows a comparison of the tractography results from SHLS-ESR and CU-GDS-COM reconstructions of a randomly selected test subject for several budgets. We can see that with as few as 5 directions, the proposed methodology is able to produce meaningful streamlines. The bottom plot in Figure~\ref{fig:tractography} (a) shows the test set average per-budget normalized Frobenius distance between the sparse and full data connectomes for both methods. We see that on average, the proposed method produces connectomes closer to the full data reconstruction than does the SHLS-ESR based framework for each sparse budget considered. To further quantify the structural connectivity matrix, we extracted and compared several graph topological metrics including network density, global efficiency, transitivity, and characteristic path length. Detailed definitions of these metrics as well as their interpretation within the context of connectome analysis can be found in \cite{Rubinov_2010}. Figure~\ref{fig:tractography} (b) displays the average and $95\%$ error bars for each of these metrics computed over the 10 subjects in the test set at different sampling budgets. We can see that the network properties of the connectomes restored using the CU-GDS-COM method converge to the full data results faster than those of the SHLS-ESR method. The SHLS-ESR generally needs $30$ or more directions to obtain a connectivity matrix that is comparable with the one obtained from 90 directions, while the CU-GDS-COM only needs around $10$. 

\subsection{Reproducibility Analysis}
We studied the ability of the proposed method to capture
meaningful variability across subjects. Due to the incorporation
of a prior in the sampling direction selection and
signal fitting, it is a valid concern that our method could be
``shrinking'' the signal estimates too much towards the ``average
brain'', especially in the sparse sample case. To investigate
this, we leveraged the test and retest data in HCP, i.e., those
 subjects that have multiple scans collected at different
times. In total, 12 non training-set subjects with their test and
retest diffusion data (24 scans) were involved in our analyses.
\par 
We compared the reproducibility (the inter- v.s. intra-subject variabilities) of the individual fODFs computed by either method. In particular, for each subject, we fit fODFs using the complete data (all 90 directions) collected at the first scan (scan-1). The data from the second scan (scan-2) was used to estimate the fODFs under a variety of budget constraints. All of the fODFs were warped to the template space to allow joint comparisons. For each budget and both reconstruction methods, within- and between-subject $\mathbb{L}^2$ errors were computed for the scan-2 fODFs using the complete data scan-1 fODFs as ground truth. The reproducibility at each voxel was summarized using the dICC: $\text{dICC}_{v} = \bar{d}_{v,bs}^2/( \bar{d}_{v,bs}^2 +  \bar{d}_{v,ws}^2)$ \citep{Zhang2017HCPg}, where $\bar{d}_{v,bs}$ and $\bar{d}_{v,ws}$ are the mean between- and within-subject errors at voxel $v$, respectively. dICC takes values ranging from 0 to 1, where a higher value indicates better reproducibility, i.e., large inter-subject variance and small intra-subject variance. In our case, a higher dICC can indicate that the priors are not dominating the estimation and that our method can properly model subject level heterogeneity from individual dMRI data. As a baseline, we also computed fODFs using just the prior mean. The per-voxel $\mathbb{L}^2$ distance between the mean fODF and scan-1 ``ground truth'' was calculated and averaged across subjects, denoted here as $\bar{d}_{v,mean}^2$.
\par
\begin{figure}
    \centering
    \begin{tabular}{c}
         \includegraphics[scale=0.5]{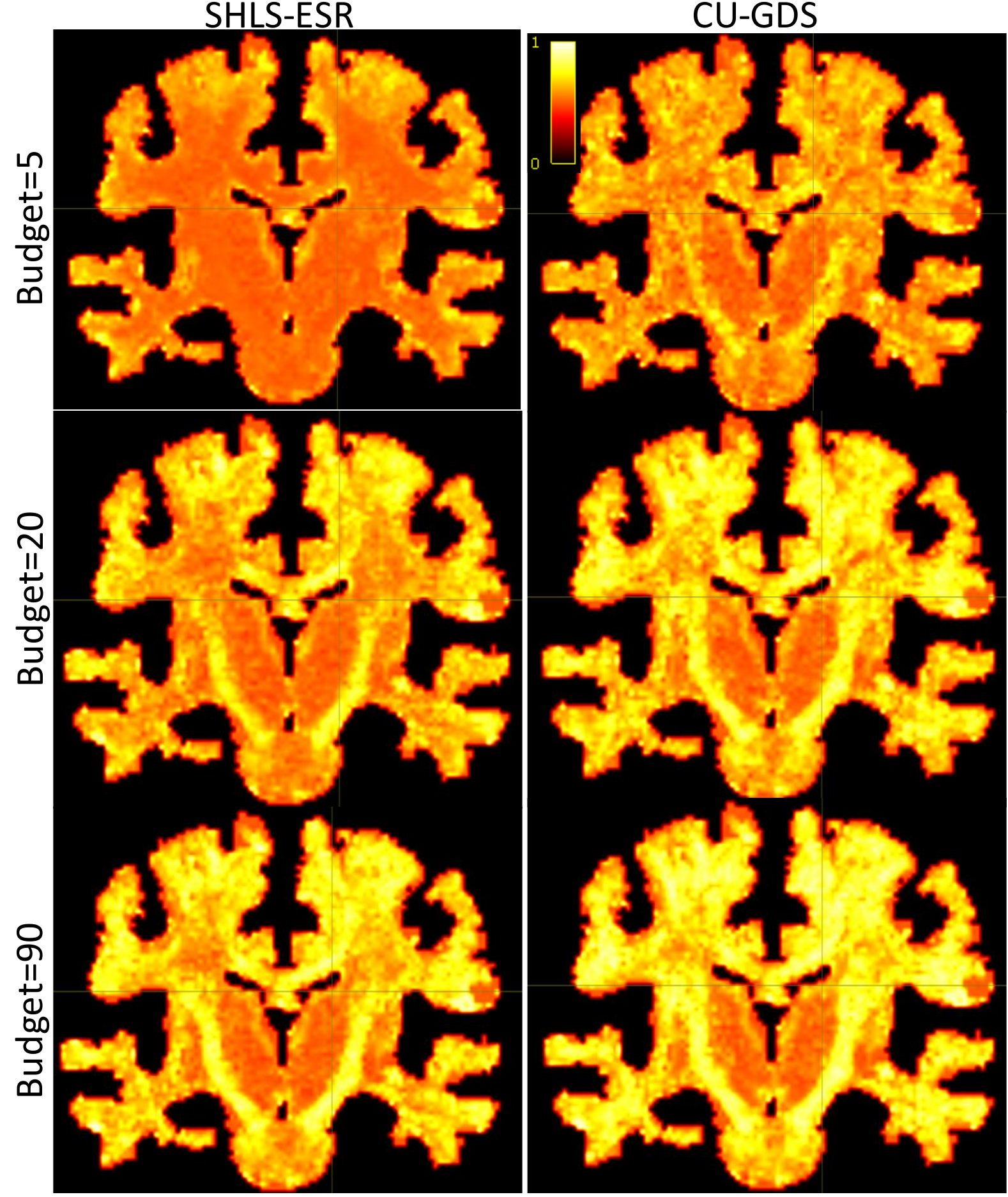} \\
         (a) \\
          \includegraphics[scale=0.25]{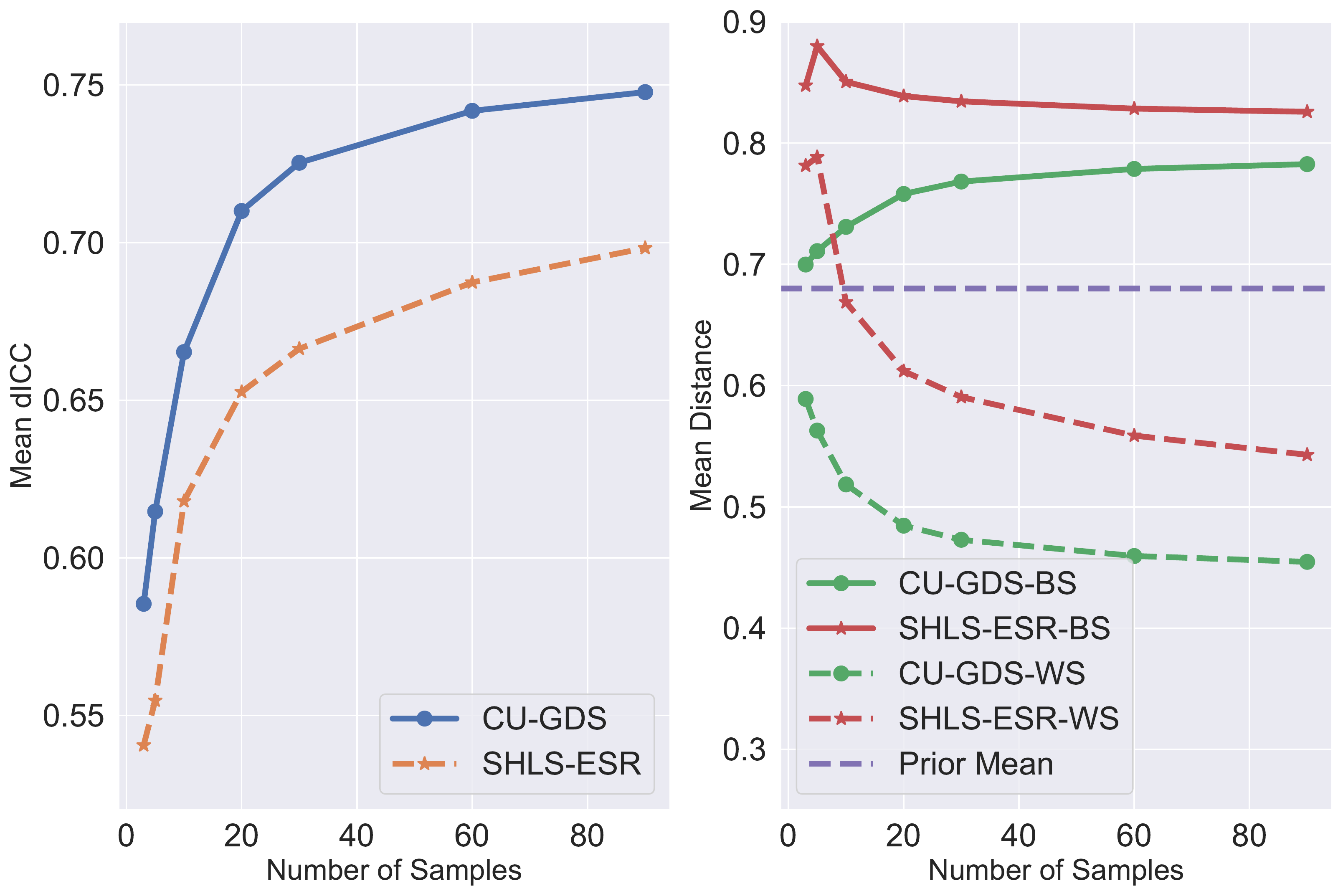} \\
          (b)
    \end{tabular}
    \caption{(a) $\text{dICC}_{v}$ using the fODFs computed from the test-retest HCP data for several budgets. (b) (Left) Averaged $\text{dICC}_{v}$. (Right) per-budget  within- (WS) and between-subject (BS) distances averaged across the white matter voxels. The horizontal line identifies the prior mean's result, i.e., $|\mathcal{V}|^{-1}\sum_{v\in\mathcal{V}} \bar{d}_{v,mean}^2$.}
    \label{fig:dICC_reproducibility}
\end{figure}

 
Figure~\ref{fig:dICC_reproducibility} (a) shows the $\text{dICC}_{v}$ at all white matter voxels in a coronal slice of the template space for several budgets. The proposed method displays better reproducibility, especially at sparse budgets, for nearly all of the voxels shown. The left panel of Figure~\ref{fig:dICC_reproducibility} (b) shows the per-budget average $\text{dICC}_{v}$, i.e. $|\mathcal{V}|^{-1}\sum_{v\in\mathcal{V}}\text{dICC}_{v}$. Our method results in better average reproducibility for all the budgets considered. The right panel of Figure~\ref{fig:dICC_reproducibility} (b) shows the per-budget within- (WS) and between-subject (BS) distances averaged across the white matter voxels, i.e. $|\mathcal{V}|^{-1}\sum_{v\in\mathcal{V}} \bar{d}_{v,ws}^2$, $|\mathcal{V}|^{-1}\sum_{v\in\mathcal{V}} \bar{d}_{v,bs}^2$, for both reconstruction methods. The prior mean result, i.e. $|\mathcal{V}|^{-1}\sum_{v\in\mathcal{V}} \bar{d}_{v,mean}^2$, is also displayed. As a Bayesian method, in the absence of much data, the estimates from CU-GDS are shrunk towards the prior, which is manifested in the relative proximity of the between- (green solid) and within-subject (green dashed) curves at low vs high budgets. Still, even for the budget=$5$ case, the proposed method exhibits significantly lower within-subject distance and higher between-subject distance compared to the prior mean (purple dashed line). Thus, CU-GDS is able to capture meaningful subject specific variability in the very sparse regime. This is in contrast to SHLS-ESR (red lines), whose performance in the budget$ < 10$ case is very poor to the point where it is inferior to the naive strategy of simply using the prior mean for each subject.
\subsection{Out of Distribution Analysis}\label{ssec:ood_analysis}
One potential disadvantage of the proposed method is its requirement of high-quality dMRI data from the population of interest. This could restrict the potential impact of the method given that such data are not always available. In this section we investigate how well the priors learned from one population generalize to another population. In particular, we used the priors estimated from the HCP young adults and applied the proposed methodology to a test dataset consisting of older adults with MCI. Since this out of distribution (OOD) data are predominantly on the $b=2,000$ shell, we used the HCP priors constructed from this shell for analysis. The FA image for each OOD subject was registered to the template space and the resulting warping function was used to map the priors into the corresponding subject spaces. CU-GDS and SHLS-ESR estimators were applied to the $b=2,000$ data for a set of increasing budgets. Again, the fODFs were constructed from the estimated diffusion signal functions by applying the FRT and spherical deconvolution from \cite{descoteaux_2009}. Due to the absence of a real ``ground truth'' for comparison, we adopt the same tactic as before and allow the SHLS-ESR reconstructed fODFs from the first 90 directions to serve this role. The fODFs constructed from the prior mean diffusion signal function were also computed. To allow direct comparison of the results to those from an ``in distribution'' (data from HCP) sample, we also computed the fODFs using CU-GDS applied to the HCP test subjects $b=2,000$ shell.
\begin{figure}
    \centering
    \includegraphics[scale=0.57]{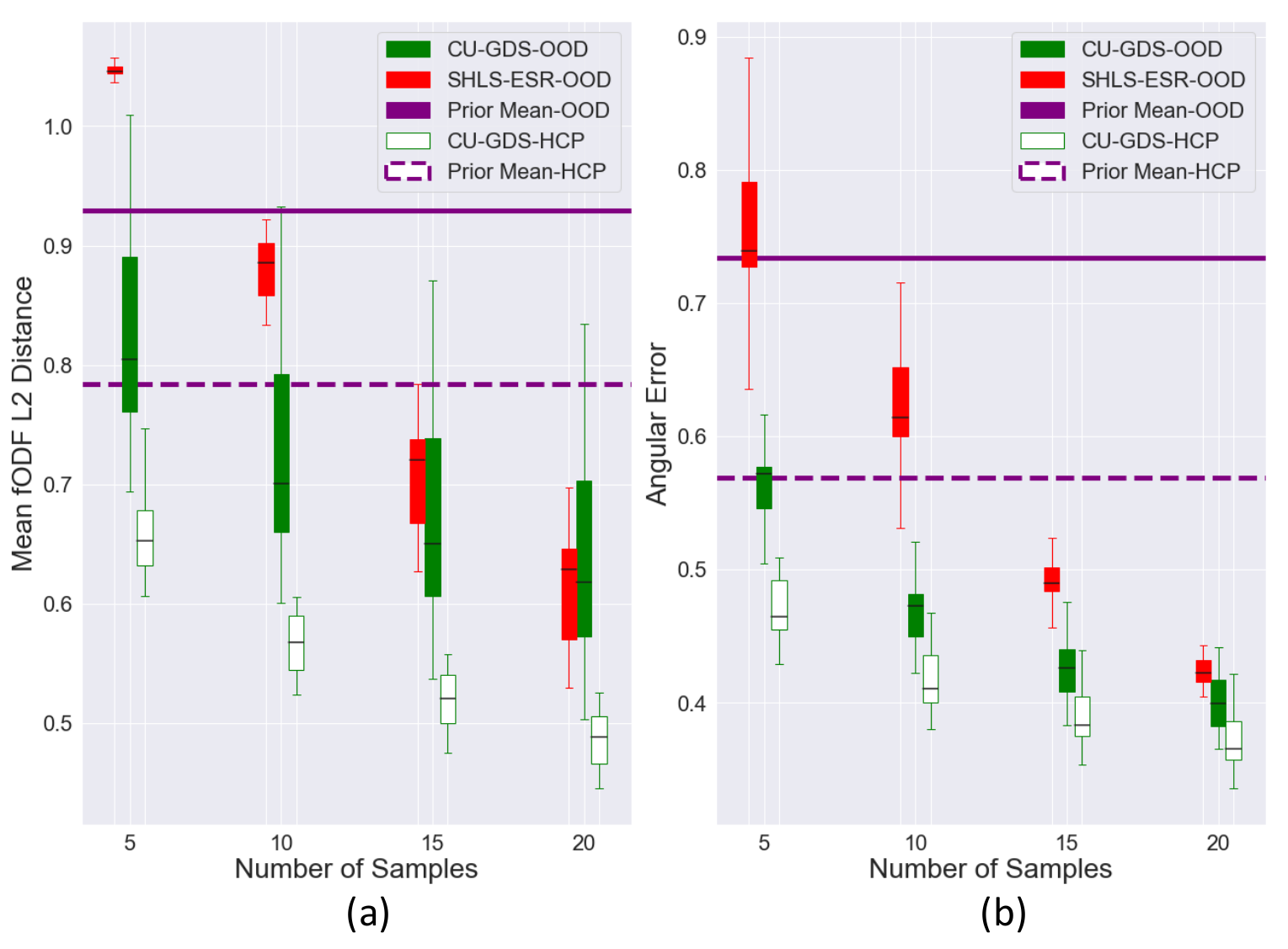}
    \caption{Average per out of distribution (OOD) subject sparse budget fODF reconstruction performance for CU-GDS (green) and SHLS-ESR (red). The median of the prior mean fODF performance is shown for reference (purple line). As measured by both $\mathbb{L}^2$ distance (a) and estimated angular error (b). For OOD data, although not as good as ``in distribution'' (the HCP case) data, CU-GDS can still better approximate the directional information computed from SHLS-ESR90 for each spares budget considered.}
    \label{fig:ood_fodfs}
\end{figure}
\par 
For each test subject, the difference between the sparse budget fODFs and high cost ground truth was measured at each voxel using both $\mathbb{L}^2$ distance and angular error. The per-subject averages of both metrics over the entire OOD sample are displayed in the box plots in Figure~\ref{fig:ood_fodfs}. For comparison, the median performance of the prior mean for both OOD and HCP samples is shown, along with the results of the CU-GDS method applied to the in distribution HCP $b=2,000$ test data. Although it does not perform as well as the in distribution case, the CU-GDS method still better approximates the SHLS-ESR90 fODFs than do those constructed from SHLS-ESR  for all of the sparse budgets considered, as measured by both $\mathbb{L}^2$ error (panel (a)) and angular error (panel (b)). Notice that even in the OOD sample, the CU-GDS method performs significantly better than the prior mean fODF reconstruction, indicating that the boost in performance is not driven by the prior mean. Comparing the in vs. out of distribution results of the CU-GDS, it is clear that access to priors learned using the same population as the test data results in better reconstructions. Still, even when trained on a historical sample from a population that is ostensibly very different than the target, our method's sparse sample estimates better approximate the full data fODFs when compared to the standard technique.

\section{Conclusion and Discussion}\label{sec:conclusion}

This paper introduces a novel methodology which incorporates high-resolution historical dMRI data to optimize diffusion weighted directions in dMRI acquisition and sparse sample diffusion signal estimation. Historical data is used to construct a prior distribution of diffusion signal at each voxel in a template space through a low-rank GP model. Given a new subject of interest for dMRI acquisition, the GP priors in the template space are mapped to the subject space and then used to define the subject-specific acquisition directions that minimize the expected integrated squared error of the proposed diffusion signal estimator. A computationally efficient greedy algorithm is proposed to select the directions and is shown to approximate the performance of the true optimal set of directions within some computable bounds. Using extensive simulation and real data experiments, we show that the combination of the proposed diffusion sampling and signal estimation generally outperforms traditional techniques. The proposed framework is significantly better in the sparse sample case, i.e, when we have limited budget in the number of diffusion directions to acquire. In applying the proposed method to analyze HCP data, we show the boost in performance of our method in diffusion signal estimation carries through to fODF reconstruction and ultimately to quality tractography estimation.
\par 
We now provide several remarks on the practical application of our method. Given the public availability of large brain imaging data sets with densely sampled high-resolution dMRI, e.g., the different types of HCP sets, the Adolescent Brain Cognitive Development (ABCD) set, and the UK Biobank set, the requirement of finding a sample data to serve in constructing the prior is easy to accomplish for many populations of interest. Even if an exact reference population cannot be acquired, results from Section~\ref{ssec:ood_analysis} indicate that, at least in some cases, our method still performs well when applying the priors learned from one population to the analysis of another. In order to construct quality empirical priors, it is important to include an adequate number of subjects in the historical training sample. In additional simulation results reported in Section S5 of the supplemental material, we found that the performance of our method was stable for $N\ge 50$. In practice, this will depend on the population of interest and, if necessary, a resampling scheme can be used to select $N$. In order to use our methodology for data acquisition of a new subject, a small number of $b=0$ scans must first be collected to provide the subject's coordinate for prior injection and  to estimate the measurement error variance $\sigma^2$. In this work, the registration between template and subject space was conducted using the FA image, but it is possible to instead use the $b=0$ image for this purpose in a real application. For estimating $\sigma^2$, although the historical sample could be used, it is important to do so for each new subject since it captures the noise introduced by the scanner, which can vary substantially between machines and scanning sites (see the supplemental material Section S2 for further discussion).
\par 
Moreover, an important question that arises in applying the proposed method to real data collection is how to correct eddy current-induced distortions and subject movements for sparsely sampled dMRI. In practice, the most popular method for motion and distortion correction is FSL's eddy toolbox, which makes several assumptions about the acquired data. For instance, it is assumed that the diffusion encoding directions span the entire sphere. In analysis not reported we observed that, for a given budget, the GDS designs had similar electrostatic energy as the ESR and typically much smaller energy than a design by random uniform sampling on $\mathbb{S}^2$. For example, notice that the GDS designs in Figure~\ref{fig:signal_error_distribution} are relatively well dispersed, even compared to the ``optimally dispersed'' ESR designs. We therefore expect the GDS designs to satisfy the dispersion requirement of FSL's eddy toolbox. Additionally, we note the current eddy toolbox in FSL uses a GP model \cite{andersson2015} to predict diffusion signal, and that this model may perform poorly for sparse data. To make it work for sparse data, one can replace the prediction equation (7) in \cite{andersson2015} with our prediction equation~\eqref{eqn:conditional_expectation_function_approximation}. Alternatively, to use the current eddy toolbox as is for the proposed acquisition method, we make the following two suggestions. 1) Acquire at least 20 directions. According to \cite{ning2015}, we expect the current eddy toolbox to work for this sparsity level. 2) Acquire blip-up-blip-down volumes for each direction. The repetition of all diffusion gradient directions using forward and reverse phase encoding provides a means to resolve signal intensity recovery in compressed areas, and gives the best distortion correction in practice \cite{brun2019}.
\par 
We conclude by offering several potential applications and extensions of the proposed framework. First, in clinical scans, long scanning times can be infeasible for a variety of reasons. The proposed method can be used to design optimal gradients to quickly acquire sparse q-space data for important clinical purposes such as surgical planning. Second, there are studies trying to increase the spatial resolution of diffusion MRI to more accurately detect white matter fiber tracts at a submillimeter isotropic resolution \cite{chang2015human}. Such spatial resolution requires long TR, and thus only very few samples can be collected in a given time period. In this sparse setting, the proposed method shows the most prominent performance boost over the standard technique. One immediate extension is to augment the statistical model to accommodate for signal attenuation in the multi-shell situation. This would allow the derivation of a joint optimization scheme for both gradient direction and magnitude in the dMRI protocol. Another extension involves multi-voxel analysis and information borrowing from neighborhood regions. The white matter tracts often group in bundles that span many voxels, inducing correlations between the signal in neighboring voxels.  Considering such correlation can increase the efficiency of the estimation procedure.
Finally, it is likely that in some situations, an appropriate historical sample for the reference population does not exist and priors from healthy subjects may have poor generalization performance, e.g. in clinical populations where particular tracts are profoundly altered due to pathology such as lesions or tumors. That said, the priors can still be of assistance in such cases, namely for identifying ``outlier regions'' where the current subject differs substantially from the reference prior. Under the Bayesian perspective, due to conjugacy we have an explicit form of the full posterior of the diffusion signal, given the observed data. In practice, this allows methods from Bayesian outlier detection, e.g. \cite{bayarri2003,zaslavsky2010}, to be adapted for our purposes. There is also a literature on outlier detection in functional data analysis that could be used in this task, e.g. \cite{gervini2012}, but we leave the specifics as an interesting direction of future inquiry.

\bibliographystyle{ieeetr}
\bibliography{bibfile,ref}

\end{document}